\pdfoutput=1
\documentclass[12pt,preprint]{aastex}
\usepackage{amsmath}
\begin{document}
\newcommand{\up}[1]{\ifmmode^{\rm #1}\else$^{\rm #1}$\fi}
\newcommand{\zdot}{\makebox[0pt][l]{.}}
\newcommand{\upd}{\up{d}}
\newcommand{\uph}{\up{h}}
\newcommand{\upm}{\up{m}}
\newcommand{\ups}{\up{s}}
\newcommand{\arcd}{\ifmmode^{\circ}\else$^{\circ}$\fi}
\newcommand{\arcm}{\ifmmode{'}\else$'$\fi}
\newcommand{\arcs}{\ifmmode{''}\else$''$\fi}

\title{Empirical calibration of the reddening maps in the Magellanic Clouds}


\author{Marek G{\'o}rski}
\affil{Universidad de Concepci{\'o}n, Departamento de Astronomia,
Casilla 160-C, Concepci{\'o}n, Chile}
\authoremail{mgorski@astro-udec.cl}

\author{Bart\l{}omiej Zgirski}
\affil{Nicolaus Copernicus Astronomical Center, Polish Academy of Sciences, Bartycka 18, 00-716, Warsaw, Poland}

\author{Grzegorz Pietrzy{\'n}ski}
\affil{Nicolaus Copernicus Astronomical Center, Polish Academy of Sciences, Bartycka 18, 00-716, Warsaw, Poland}
\affil{Universidad de Concepci{\'o}n, Departamento de Astronomia,
Casilla 160-C, Concepci{\'o}n, Chile}

\author{Wolfgang Gieren}
\affil{Universidad de Concepci{\'o}n, Departamento de Astronomia,
Casilla 160-C, Concepci{\'o}n, Chile}

\author{Piotr Wielg{\'o}rski}
\affil{Nicolaus Copernicus Astronomical Center, Polish Academy of Sciences, Bartycka 18, 00-716, Warsaw, Poland}

\author{Dariusz Graczyk}   
\affil{Nicolaus Copernicus Astronomical Center, Polish Academy of Sciences, Bartycka 18, 00-716, Warsaw, Poland}
\affil{Universidad de Concepci{\'o}n, Departamento de Astronomia,
Casilla 160-C, Concepci{\'o}n, Chile}

\author{Rolf-Peter Kudritzki}   
\affil{Institute for Astronomy, Honolulu, HI, USA.}
\affil{Munich University Observatory, Munich, Germany. }

\author{Bogumi\l{} Pilecki}   
\affil{Nicolaus Copernicus Astronomical Center, Polish Academy of Sciences, Bartycka 18, 00-716, Warsaw, Poland}
\affil{Universidad de Concepci{\'o}n, Departamento de Astronomia,
Casilla 160-C,Concepci{\'o}n, Chile}

\author{Weronika Narloch }
\affil{Universidad de Concepci{\'o}n, Departamento de Astronomia, Casilla 160-C, Concepci{\'o}n, Chile}

\author{Paulina Karczmarek}
\affil{Warsaw University Observatory, Al. Ujazdowskie 4, 00-478, Warsaw,  
Poland}
\affil{Universidad de Concepci{\'o}n, Departamento de Astronomia, Casilla 160-C, Concepci{\'o}n, Chile}

\author{Ksenia Suchomska}  
\affil{Nicolaus Copernicus Astronomical Center, Polish Academy of Sciences, Bartycka 18, 00-716, Warsaw, Poland}

\author{M{\'o}nica Taormina}   
\affil{Nicolaus Copernicus Astronomical Center, Polish Academy of Sciences, Bartycka 18, 00-716, Warsaw, Poland}


\begin{abstract}
We present reddening maps of the Large Magellanic Cloud (LMC) and Small Magellanic Cloud (SMC), based on color measurements of the red clump. Reddening values of our maps were obtained by calculating the difference of the observed and intrinsic color of the red clump in both galaxies. To obtain the intrinsic color of the red clump, we used reddenings obtained from late-type eclipsing binary systems, measurements for blue supergiants and reddenings derived from Str\"{o}mgren  photometry of B-type stars. We obtained intrinsic color of the red clump $(V-I)_0$ =  0.838 $\pm$ 0.034 mag in the LMC, and $(V-I)_{0}$ =  0.814 $\pm$ 0.034 mag in the SMC. We prepared our map with 3 arcmin resolution, covering the central part of the LMC and SMC.
The mean value of the reddening is E$(B-V)_{\mathrm{LMC}}$=0.127 mag and E$(B-V)_{\mathrm{SMC}}$=0.084 mag for the LMC and SMC, respectively. The systematic uncertainty of the average reddening value assigned to each field of our maps is 0.013 mag for both Magellanic Clouds. Our reddening values are on average higher by 0.061 mag for the LMC and 0.054 mag for the SMC, compared with the maps of Haschke et al. (2011). We also compared our values with different types of reddening tracers. Cepheids, RR Lyrae stars, early-type eclipsing binaries and other reddening estimations based on the red clump color on average show reddenings consistent with our map to within a few hundredths of magnitude.
\end{abstract}



\section{Introduction}
The Large Magellanic Cloud (LMC) and Small Magellanic Cloud (SMC) are dwarf galaxies, the most massive satellites of the Milky Way. Because of their proximity and stellar populations, they are important objects to study a large variety of astrophysical phenomena. Both galaxies have been the targets of numerous optical surveys, 
like the Massive Compact Halo Objects (MACHO) survey (Alcock et al. 2000), the Optical Gravitational Lensing Experiment (OGLE, Udalski et al. 2008a,b) or the VISTA Magellanic Cloud (VMC) survey (Cioni et al. 2011). LMC plays a crucial role for the distance scale, as the distance to the LMC is used to calibrate the zero-point of the brightness of different standard candles, like Cepheids (Macri et al. 2015) or the tip of the red giant branch (G{\'o}rski et al. 2018; Madore et al. 2018). Those calibrations play a fundamental role in 
the Hubble Constant value determination (Freedman et al. 2001; Freedman et al. 2019; Riess et al. 2016; Riess et al. 2019; Yuan et al. 2019). Both galaxies are also the subject of stellar populations and 3D structure studies 
(Ripepi et al. 2012; Jacyszyn-Dobrzeniecka et al. 2017). In recent years both Magellanic Clouds were the subject of intense scrutiny of the Araucaria Project (Gieren et al. 2005), which led to a precise distance determination for both galaxies (Pietrzy{\'n}ski et al. 2013; Graczyk et al. 2014; Pietrzy{\'n}ski et al. 2019), a determination of the Cepheid period-luminosity relation dependence on metallicity  ( Wielg{\'o}rski et al. 2017; Gieren et al. 2018), the determination of precise fundamental physical parameters of eclipsing binary systems containing red giants (Graczyk et al. 2018) and a calibration of a new method to use blue supergiant stars as extragalactic distance indicators (Urbaneja et al. 2017).

For all these studies it is important to properly take into account the effect of interstellar extinction, which is caused by dust located in the Milky Way and within the LMC or SMC. 
Since extinction is dependent on the wavelength (A$_{\lambda}$), it changes the observed colors of the stars by shifting the color toward red. The difference of the true and observed colors of the stars is known as the color excess or reddening E$(\lambda_1 - \lambda_2)=A_{\lambda_1} - A_{\lambda_2}$. The connection of the extinction in the visual band $A_V$ and the color excess E$(B-V)$ is known as the reddening law.  In 1989 Cardelli et al. found that the reddening law can be approximated with just one parameter, $R_V$, which is defined as 
the ratio of total-to-selective extinction R$_V$=A$_V$/E$(B-V)$. 

Because in different regions of the LMC and SMC we can expect different dust contents, different distributions of dust grain sizes and chemical composition, the reddening and reddening law 
can vary within both galaxies. However, a number of studies have shown that for the diffuse interstellar medium for the optical and near-infrared part of the spectrum, extinction curves have a similar shape and the mean value of the reddening law is R$_V$=3.1 (Cardelli et al. 1989; Pei 1992; Weingartner \& Drain 2001; Fitzpatrick \& Massa 2007;  Schlafly \& Finkbeiner 2011).

Measurements of the reddening are performed with various techniques. Zaritsky (1999) obtained reddening from 
fitting the effective temperature, the luminosity and the extinction to the four-filter photometry of a large number of stars in the LMC. 
Pejcha \& Stanek (2009) measured difference of the observed and intrinsic colors 
of RR Lyrae stars. Inno et al. (2016) used multiwavelength apparent distance moduli to fit reddening and true distance for individual Cepheids in the LMC disk.
Larsen et al. (2000) measured individual reddenings for B-type stars in the LMC and SMC based on the Str\"{o}mgren \textit{uvby} photometry. 
Urbaneja et al. (2017) determined reddening, extinction, and R$_V$ values for a large number of blue supergiant stars in the LMC. 
Taormina et al. (2019), Bonanos et al. (2011) and Groenewegen \& Salaris (2001) obtained reddening for early-type eclipsing binary systems in the LMC. 

One of the most important methods used to measure reddening is based on the red clump color excess, that is the difference of the observed and unreddened, intrinsic mean color of the red clump stars E$(V-I)=(V-I)-(V-I)_0$. 
Udalski et al. (1999) measured red clump color excess in both Magellanic Clouds based on the OGLE II photometry. To obtain red clump $(V-I)_0$ intrinsic color in the LMC, they used 
reddening measurements for the NGC 1850 and NGC 1835 stellar clusters, and reddening determined to HV 2274 eclipsing binary system. Olsen \& Salyk (2002) used red clump stars to obtain differential reddening to 50 fields located in the LMC disk. 
In this case the intrinsic color $(V-I)_0=0.92$ mag was adopted, to match the mean value of Schlegel et al. (1998) Galactic reddening map toward the LMC. We have to note that Galactic reddening toward LMC was not measured, but interpolated from surrounding regions on the sky (Schlegel et al. 1998). 
If this value is used to obtain total reddening toward objects in the LMC, the reddening is systematically underestimated by the value corresponding 
to the intrinsic dust content in the LMC. The $(V-I)_0=0.92$ mag intrinsic red clump color was used also by Subramaniam (2005) and Haschke et al. (2011). Both reddening maps, 
as a consequence, show systematically lower values of reddening, compared to Udalski et al. (1999) reddening maps. 
Usually the correction to obtain the correct value of the reddening for the Magellanic Clouds is achieved by applying a small offset, that has to be added 
to the values presented in the Haschke et al. (2011) maps (Pietrzy{\'n}ski et al. 2013; Graczyk et al. 2014, 2018; Gieren et al. 2018). 
These offset values are usually based on a comparison with other measurements of the reddening for specific stars or fields. 
Not including this correction result in a systematic error of numerous studies. Underestimation of the reddening as a consequence can lead to underestimation of the 
absolute magnitude of standard candles.


Tatton et al. (2013) prepared a detailed reddening map of the central region of the LMC, containing 30 Doradus, based on the E$(J-K)$ red clump color excess. The intrinsic $(J-K)_0$ color was obtained with the star formation history. Choi et al. (2018) measured E$(g-i)$ color excess of the red clump stars covering 165 deg$^2$ of the LMC disk.  To obtain the intrinsic $(g-i)_0$ color of the red clump, they measured color in the outer regions, assuming negligible internal dust contribution in those fields. The $(g-i)_0$ color was derived with E$(B-V)=0.065$ mag
reddening originating from Schlegel et al. (1998) Galaxy reddening maps. Authors used three different values of the intrinsic color of the red clump, depending on the distance from LMC center. The change of the intrinsic color of the red clump is interpreted as a result of different metallicity and age of the red clump stars in the central regions and in the outer regions. The $(g-i)_0$ value, that was used to obtain color excess in the LMC bar, was however derived from fields located more than 2 deg from the center of the LMC.

In this paper we decided to prepare reddening maps of the central regions of the LMC and SMC based on the red clump color, with $(V-I)_0$ intrinsic color derived from 
various reddening tracers. Potential discrepancies of the $(V-I)_0$ intrinsic color, obtained for different methods can reveal systematic differences of the reddening affecting different type of stars. 

Our paper is organized as follows. In section 2 we present our previous reddening measurements, and measurements available in the literature. 
For each of the objects with known reddening we calculate apparent $(V-I)$ color of the red clump, and use both values to obtain intrinsic, unreddened $(V-I)_0$ color of the red clump separately in both Magellanic Clouds. 
In subsection 2.2 we calculate red clump E$(V-I)$ color excess for a number of fields covering both Magellanic Clouds. In section 3 we discuss our results and uncertainty of our measurements. We compare our maps with other measurements and discuss internal dust contribution and tracer dependant reddening. We also discuss $(V-I)_0$ intrinsic color of the red clump stars, its variation over the LMC plane and the impact of the reddening law on our maps. Finally, in section 4 we summarize our results.


\section{Reddening measurements and red clump color measurements} 
To calculate the intrinsic $(V-I)_0$ color of the red clump in the LMC and SMC, we selected a number of reddening measurements to single stars or fields in both galaxies. 
We used three different reddening tracers: late-type eclipsing binaries, blue supergiants, and B-type stars. 
Other measurements mentioned in the introduction were used to compare reddening values, and investigate potential dependence of the reddening on different types of tracers. 
Recently, Graczyk et al. (2018) published a catalog of fundamental physical parameters of 20 late-type eclipsing binaries in the LMC. The authors were able to measure the interstellar extinction in the direction to each of the target systems with two methods. The first method is based on the comparison of the effective temperatures derived from the atmospheric analysis and observed $(V-I)$ and $(V-K)$ colors of each system. The color excess is interpreted as the total reddening toward those stars. 
The second method uses a calibration of the equivalent width of the interstellar absorption Na I D1 line and a gas quantity in the line of sight with a gas-to-dust ratio to estimate reddening (Munari \& Zwitter 1997). 
In principal, the Magellanic Cloud intrinsic gas content and Galactic component are separated in the velocity space. Those measurements were performed in selected orbital phases of the systems, to ensure 
sufficient separation of the galactic lines from stellar lines in the spectrum. The same technique was used previously 
(Graczyk et al. 2014) to obtain reddening values toward four eclipsing systems in the SMC. The coordinates and reddening values toward analyzed systems for both Magellanic Clouds are reported in Table 1.

Urbaneja et al. (2017) analyzed high quality spectra for nearly 90 blue supergiant stars in the LMC from which they were able to obtain the visual extinction (A$_V$), reddening E$(B-V)$ and reddening law R$_V$. 
They compared predicted spectral energy distribution with the observed one, based on the observed photometric colors and magnitudes. The authors decided to 
report only line-of-sight values, and do not consider separated contributions for the Milky Way and the LMC. E$(V-I)$ color excess obtained from measurements obtained 
by Urbaneja et al. (2017) are reported in Table 2. Schiller (2010) used similar technique to obtain E$(B-V)$ color excess for 31 objects in the SMC. In this case, however, the reddening was obtained with fixed reddening law R$_V$ = 3.1. Reddenings obtained for objects in the SMC are listed in Table 3.

Larsen et al. (2000) measured individual reddenings for B-type stars in the LMC and SMC. The reddening measurements are based on the ($b-y$) vs. c1 indices calculated from the Str\"{o}mgren \textit{uvby} photometry. The E$(b-y)$ color excess is transformed to the E$(B-V)$ reddening with the E$(B-V)=1.4 \times$ E$(b-y)$ formula. Reddening values obtained by Larsen et al. (2000) are summarized in Table 4. Figures 1 and 2 present location of the reddening tracers in the LMC and SMC, reported in this section. 


\subsection{Intrinsic color of the red clump} 

To obtain $(V-I)$ color of the red clump stars, we selected from the OGLE-III photometric maps (Udalski et al. 2008a,b) stars in the 
5 $\times$ 5 arcmin fields centered on the reddening tracers. In the next step we constructed color-magnitude diagram, and selected red clump stars based on the $(V-I)$ color 
and I-band magnitude. In each case, the selection was done manually, but usually the red clump is enclosed in the  0.5 $< (V-I) <$ 1.6 color and 17 $<$ I $<$ 19.5 magnitude range. The average photometric uncertainty for the red clump stars is 0.03 mag for V- and I- bands. Figure 3 presents exemplary I-band vs. $(V-I)$ color-magnitude diagram of stars in the 5 $\times$ 5 arcmin field centered on system LMC-ECL-06575. For each field we prepared a histogram of the $(V-I)$ colors of selected stars. Finally, with the least squares technique we fit function (1) to the histogram of the $(V-I)$ color,  

\begin{equation}
n(k)=a+b(k-k_{RC})+c(k-k_{RC})^2+\frac{N_{RC}}{\sigma_{RC}\sqrt{2\pi}}exp[-\frac{(k-k_{RC})^2}{2\sigma^2_{RC}}]\mathrm.
\label{eq:RC}
\end{equation}

Fitted function is composed of a Gaussian representing the red clump stars and a second-order polynomial approximating the stellar background (e.g. RGB stars). The $n(k)$ is the number of stars in each $(V-I)$ histogram bin of the $k$ color. The $\sigma_{RC}$ is the spread of red clump stars colors, N$_{RC}$ corresponds to the number of red clump stars in the histogram and $k_{\mathrm{RC}}$ parameter is the mean color of the red clump stars. Measured mean colors of the red clump are reported in Tables 1, 2, 3 and 4. 
Exemplary histogram of the red clump stars in the field centered on the LMC-ECL-06575 system with fitted function (1)  is presented on Figure 4.
For each field with measured $(V-I)_\mathrm{RC}$ color of the red clump we calculated intrinsic, unreddened $(V-I)_0$ color of the red clump, by 
applying reddening measured with the reddening tracer located in this field. Following total-to-selective extinction for the Landolt V- and I- bands based on Cardelli et al. (1989) parametrization, we used $(V-I)_0=(V-I)_{RC} - 1.318 \times$ E$(B-V)$ formula. For the blue supergiants in the LMC, the procedure was slightly different. From the total number of 90 objects presented by Urbaneja et al. (2017) we measured red clump color only for 33 objects. We rejected a number of objects, which were located in a region with high reddening values or were outside the coverage of the OGLE-III maps. For each of the remaining objects, we decided to calculate E$(V-I)$ color excess directly from the monochromatic E(4405-5495) color excess with the R$_{5495}$ reddening law values, 
obtained by Urbaneja et al. (2017). Calculated values of the E$(V-I)$ are reported in Table 2. The R$_{5495}$ and E(4405-5495) values were obtained thanks to the courtesy of the authors. In Tables 5 and 6, we present the mean $(V-I)_0$ unreddened color of the red clump, calculated separately for each method, for both Magellanic Clouds. We also report the spread of calculated intrinsic color of the red clump within each method in Tables 5 and 6.

Values of the intrinsic color of the red clump stars obtained with different reddening tracers are consistent with each other to within a few hundredths of magnitude. The differences can be fully attributed to systematic errors of reddening determination of each method. Therefore, we 
decided to adopt the mean of the values obtained for each method as the intrinsic $(V-I)_0$ color of the red clump. 

$\mathrm{LMC:~ (V-I)_{0} =  0.838~mag ~(std = 0.017 ~mag) }$

For the SMC we calculated the intrinsic color of the red clump as the mean of the $(V-I)_0$ colors obtained for each method, which are presented in Table 6.

$\mathrm{SMC:~ (V-I)_{0} =  0.814~mag ~(std = 0.015 ~mag)}$

We note that $(V-I)_{0}$ intrinsic colors were obtained with the reddening law R$_V$=3.1. 
The reported spread (std) is calculated as the standard deviation of the intrinsic colors obtained with different methods.
 
\subsection{Reddening Maps}
Our reddening maps are based on the $(V-I)$ red clump color measured in 3 $\times$ 3 arcmin fields covering central regions of the LMC and SMC. 
For each field we selected red clump stars from the color-magnitude diagram. In the first step, we selected stars of I- band magnitude in the range of 
16.5 to 19.5 mag and $(V-I)$ color from 0.5 to 1.5 mag. In the next step we fitted function (1) to the histogram of the color. We obtained the mean $(V-I)_{RC}$ color of the red clump with corresponding uncertainty, spread of the red clump stars colors ($\sigma_{RC}$) and number of red clump stars in the field (N$_{RC}$). Based on the calculated mean color and spread of the color of red clump stars, selection box was slightly adjusted. Following Choi et al. (2019), we decided to use polygon (tilted box - e.g. Figure 3) for magnitude and color selection. The selection boundary roughly follows the reddening vector, and secures smaller contamination of selected stars with background red giant branch stars. Then the fitting procedure was repeated. The correction for each field was smaller than 0.02 mag. For the SMC, the initial magnitude selection range was shifted by 0.5 mag.

To obtain the E$(V-I)$ color excess we calculated the difference of the apparent $(V-I)_{RC}$ color of the red clump and the adopted intrinsic $(V-I)_0$ color, separately for both Magellanic Clouds. We used Cardelli et al. (1989) R$_V$=3.1 reddening law to calculate E$(B-V)$=E$(V-I)$/1.318. Figures 5 and 6 present reddening maps calculated for the LMC and SMC, with the 3 arcmin resolution. Table 7 and 8 show the format of our maps. For a number of fields located in the outer part of our maps, the number of red clump stars was below 30. We decided to reject those fields, since low number of red clump stars can result in a systematic error of the derived parameters. Reddening in those regions can be obtained from maps with lower spatial resolution, available online with the 3 arcmin resolution maps ( https://araucaria.camk.edu.pl/index.php/magellanic-clouds-extinction-maps ).

\section{Discussion}

Our 3 arcmin resolution maps (Figure 5) show the clumpy and filamentous nature of the reddening in the LMC. The minimal reddening obtained for the LMC is E$(B-V)=0.05$ mag and can be interpreted as the Galactic reddening toward LMC. The mean value of the reddening is E$(B-V)=0.127$ mag and reaches maximum E$(B-V)=0.4$ mag in the fields of star forming regions close to 30 Doradus. Another region with slightly higher reddening values is present for fields between RA 04$^h$30$^m$ and  05$^h$00$^m$, extending south from 68$^{\circ}$ declination. This particular feature will be discussed later in this section. The median number of red clump stars in fields is 164, and reaches more than 1500 stars in the LMC bar. Median uncertainty of the calculated mean color of the red clump is 0.003 mag, and for 97\% of fields is below 0.01 mag.

For the SMC, the minimal reddening is E$(B-V)=0.04$ mag, and mean reddening is E$(B-V)=0.08$ mag. The maximum value is E$(B-V)=0.227$ mag. The median number of red clump stars in the SMC is 116 with maximum 1000 in the central part. Median uncertainty of the calculated mean color of the red clump is 0.03 mag, and for 95\% of fields is below 0.01 mag.

As the systematic uncertainty of the mean reddening values in each field of our maps we adopt the spread, calculated as the standard deviation of average values of the $(V-I)_0$ colors obtained for each method, that is 0.016 mag for the LMC, and 0.015 mag for the SMC. With Cardelli et al. (1989) R$_V$=3.1 reddening law we obtain an error of the E$(B-V)$ reddening 0.013 mag for the LMC. For the SMC we obtain 0.011 mag, but conservatively we adopt 0.013 mag as the uncertainty of the mean reddening. The adopted systematic uncertainty of our maps does not include the potential variation of the red clump color within LMC or SMC. This problem is discussed in the following subsection. 

\subsection{Internal reddening  in the LMC}
In 2009 Subramanian \& Subramaniam used the spread of the I-band magnitude and color of the red clump stars to calculate internal reddening in the LMC. Dust located in front of the LMC should change only the mean color of the red clump. Dust located in the LMC, changes mean color, but also increases the spread of the color of red clump stars. 
Figure 7 and 8 show a map of the color spread of the red clump stars $\sigma_{RC}$. Regions with the high values of $\sigma_{RC}$ can be identified as the regions of high reddening values in both Magellanic Clouds. However, our attention draws that slightly higher reddening values in the LMC, visible for fields between RA 04$^h$30$^m$ and  05$^h$00$^m$, extending south from 68$^{\circ}$ declination is not visible in this case. It suggest that this higher value of the reddening is caused by dust located in front of the LMC, possibly in our Galaxy, which was also discussed by Haschke et al. (2011).

\subsection{Comparison with other reddening determinations}
We compared our maps with different reddening measurements, obtained for the LMC and SMC.  Bonanos et al. (2011) measured reddening toward the O-type massive eclipsing binary system LMC-SC1-105. The reddening and reddening law were obtained by computing the $(B-V)_0$ intrinsic color of the system from the atmosphere model, and fitting optical and near-infrared photometry with the Cardelli et al. (1989) reddening law parametrization. While the best fit has reddening law value R$_V$=5.8 with corresponding reddening E$(B-V)= 0.11 \pm 0.01$ mag, authors provide also reddening value E$(B-V)=0.18$ mag for fixed R$_V$=3.1. From our maps we obtain reddening value E$(B-V)=0.141$ mag. Groenewegen \& Salaris (2001) based on the UV/optical spectrum and optical data obtained reddening toward the HV 2274 eclipsing binary system in the LMC, E$(B-V) = 0.110$ mag. Reddening value calculated from our maps is E$(B-V)=0.119$ mag, and agrees very well with this measurement. Taormina et al. (2019) analyzed two early-type eclipsing binaries in the LMC:  BLMC01 and BLMC02. For both systems reddening was measured with the Na I sodium line. 
Authors obtained reddening E$(B-V) = 0.187 $ mag for the BLMC01 and 0.088 mag for BLMC02. From our reddening maps we obtain  E$(B-V) = 0.211 $ mag for BLMC01 and 0.0887 mag for BLMC02. We also compare our maps with results obtained for the NGC 1850 and NGC 1835 star clusters (Walker 1993; Lee 1995). For the NGC 1850 cluster, reddening E$(B-V)=0.15 $ mag was obtained with the \textit{UBV} photometry. Our maps provide E$(B-V) =  0.14$ mag. For the NGC 1835 cluster, reddening E$(B-V)=0.13$ mag was obtained. Reddening obtained with our maps is E$(B-V)=0.12$ mag.

Another comparison is made with the Inno et al. (2016) reddenings estimated for individual Cepheids. From the total of 3924 Cepheids  we were able to calculate reddening for 1133 objects from our maps. We selected fundamental pulsators with V- band magnitude brighter than 18. The mean difference between our values and reddening values of Inno et al. (2016) is  $<($E$(B-V)_{G19}-$E$(B-V)_{I16})>=0.004$ mag with a standard deviation of 0.087 mag, where E$(B-V)_{G19}$ are reddenings calculated in this paper, and E$(B-V)_{I16}$ are the values calculated by Inno et al. (2016). Figure 9 shows histogram of the reddening difference. The 0.087 mag spread of the reddening difference can be caused by large errors of reddening measurements for single Cepheids, spread of the reddenings obtained to particular Cepheids and large differences of the reddening law R$_V$ values for particular stars. 
We note that Inno et al. (2016) obtained reddening value with the R$_V$ value fixed to 3.2. For the first overtone Cepheids (338 objects) the mean difference of our values and Inno et al. (2016) values is 0.06 mag. Since we can not find any reason why there should be reddening difference for fundamental mode Cepheids and first overtone stars, we decided to exclude the latter ones from the comparison. Figure 10 presents spatial distribution of the difference over the LMC. 

Pejcha \& Stanek (2009) used period-color relations to determine extinction for sample of 9393 RR Lyrae stars in the LMC.
The mean difference between our values and reddening values obtained by Pejcha \& Stanek (2009) is  $\langle($E$(B-V)_{G19}-$E$(B-V)_{P09})\rangle=-0.024$ mag with a standard deviation of 0.043 mag, where E$(B-V)_{G19}$ are reddenings calculated in this paper, and E$(B-V)_{P09}$ are the values calculated by Pejcha \& Stanek (2009). 
Figure 11 shows histogram of the reddening difference. The offset visible on the histogram can be attributed to the systematic uncertainty of the intrinsic red clump $(V-I)_0$ color adopted in this paper, or systematic error of the RR Lyrae intrinsic $(V-I)_0$ color error, as the rms scatter of the color-period relation reported by the authors is 0.065 mag. Figure 12 presents spatial distribution of the difference over the LMC.

We also compared our maps with Haschke et al. (2011) reddening maps for the LMC.
For each field of Haschke et al. (2011) maps we calculated reddening. The field size was matching the size used by the authors. Since  
Haschke et al. (2011) results are reported as the E$(V-I)$ color excess, we transformed this values to E$(B-V)$ = E$(V-I)$/1.318 with the Cardelli et al. (1989) R$_V$=3.1 reddening law. 
The mean difference between our values and reddening values of Haschke et al. (2011) for the LMC is  $\langle($E$(B-V)_{G19}-$E$(B-V)_{H11})\rangle=0.061$ mag with a standard deviation of 0.012 mag, where E$(B-V)_{G19}$ are reddenings calculated in this paper, and E$(B-V)_{H11}$ are the values calculated by Haschke et al. (2011). Histogram of the difference between our values and reddening values of Haschke et al. (2011) is presented in Figure 13. Figure 14 presents spatial distribution of the difference over the LMC. 
For the SMC we obtain offset $\langle($E$(B-V)_{G19}-$E$(B-V)_{H11})\rangle=0.061$ mag with a standard deviation of 0.006 mag. Since the reddening measurement method in this paper is very similar to measurements in the paper of Haschke et al. (2011), small value of the spread of the difference is expected, and is caused by different red clump star selection criteria, and different numerical approach used by Haschke et al. (2011). The 0.061 mag offset is caused by different intrinsic color of the red clump adopted by Haschke et al. (2011), and color calculated in this paper. 

Our final comparison is done with the reddening maps based on the red clump E$(g-i)$ color excess, prepared by Choi et al. 2019. The mean difference between our values and reddening values obtained by Choi et al. (2019) for the LMC is $\langle($E$(B-V)_{G19}-$E$(B-V)_{C19})\rangle=0.008$ mag with a standard deviation of 0.029 mag, where E$(B-V)_{G19}$ are reddenings calculated in this paper, and E$(B-V)_{C19}$ are the values obtained by Choi et al. (2019). Histogram of the difference is presented in Figure 15. Figure 16 presents spatial distribution of the difference over the LMC.

\subsection{Red clump color intrinsic color}

One of the results presented in this paper is the red clump intrinsic color: $(V-I)_0=0.838$ mag for the LMC and 0.815 mag for the SMC. The uncertainty of both values is 0.034 mag, including the photometric uncertainty. Those values are significantly different from the value $(V-I)_0=0.92$ mag adopted by Olsen \& Salyk (2002), Subramaniam (2005) and Haschke et al. (2011) for the LMC, and $(V-I)_0=0.89$ mag adopted by Haschke et al. (2011) for the SMC. The $(V-I)_0=0.92$ mag was however not measured, but fixed to match the mean reddening toward LMC from Schlegel et al. (1998) Galactic reddening map. Since the red clump mean color depends on age and metallicity, the lower value of the color means that the red clump is either younger or more metal poor. Subramaniam (2005) discussed that lowering of the zero-point from 0.92 to 0.82 would mean that the red clump stars have a metallicity around Z=0.001 for adopted age. Pawlak (2016) measured LMC red clump color $(V-I)_0 = 0.87$ mag, by correcting red clump color in the OGLE IV fields located in the outskirts of the LMC with the Schlegel et al. (1998) reddening. Agreement of our maps with the maps of Choi et al. (2019), also support $(V-I)_0=0.838$ mag red clump intrinsic color in the central part of the LMC. 

However, since the LMC bar red clump population should be younger by 2 Gyr compared to disk population, and a shallow metallicity gradient is observed in the LMC, we can expect small variation of the red clump intrinsic color over the galaxy. Choi et al. (2019) adopted different values of the $(g-i)_0$ red clump intrinsic color, based on distance from the galaxy center. However, in the regions covered in our maps, only one, constant value of the intrinsic color was used. Based on Girardi \& Salaris (2001) synthetic population models, we can estimate that color variation should be smaller than 0.01 mag. For example, from Table 1 in Girardi \& Salaris (2001) paper, for adopted metallicity Z$=$0.001, for age $=$ 6 Gyr we obtain $(V-I)_0 = 0.826$ mag, and for age $=$ 4 Gyr we obtain $(V-I)_0 = 0.828$ mag. Additionally, we can expect that for composite population of red clump stars, observed in the LMC, mean color of the red clump stars should be less sensitive to age and metallicity.


The visual inspection of the reddening differences of various reddening tracers with respect to our maps (Figures 10, 12, and 17), does not reveal any significant trend. To investigate this in more detail, we selected tracers located over the LMC bar. The average difference $\langle$E$(B-V)_{G19}-$E$(B-V)_{T}\rangle$ for tracers inside the dashed box (Figure 17) is shifted by 0.005 mag in comparison to average value of all tracers, with the uncertainty of this value 0.006 mag. Similar selection for difference map based on the 
RR Lyrae stars (Pejcha \& Stanek 2009 - Figure 12) neither show significant difference. For the Cepheid reddening difference map (Figure 10), we calculate -0.015 mag offset. While this small offset can be arising from the systematic error of the reddening derived for the Cepheids, we conservatively estimate this value as the upper constraint for intrinsic red clump color variation caused by age and metallicity differences of the red clump stars.

\subsection{Tracer dependent reddening}
In order to obtain $(V-I)_0$ intrinsic color of the red clump, that works as the zero-point of our maps, we used different types of reddening tracers. However, different types of reddening tracers can be affected by the reddening  in a different way. Indeed, Zaritsky (1999) found that visual extinctions for stars with effective temperatures between 5500 and 6500 K is smaller by several tenths of a magnitude than for stars with temperatures greater than 12,000 K in the LMC. The reason for those differences can be the nature of the dust and gas itself - we can expect different dust contents, different distributions of dust grain sizes and chemical composition surrounding different types of stars. Another reason can arise from complex structure of tracer hosting galaxy. If different types of stars are not colocated, we should observe systematic difference of reddening. Tracer located in front of observed galaxy, will be affected by the reddening less, than those located further. Finally, differences can be caused by systematic errors of reddening determination. We see this effect 
for late-type eclipsing binaries. Reddenings E$(B-V)$ derived with the sodium Na I line is systematically lower by 0.02 mag compared to reddenings derived from the effective temperature.

Despite the fact that our adopted intrinsic color of the red clump is based on late- and early-type stars, the maximum difference between those tracers transforms to 0.02 mag difference 
of the E$(B-V)$ reddening. The calculated RMS for all type of tracers used to obtain intrinsic color of the red clump is 0.012 mag, and can be fully attributed to systematic errors of reddening determination. Further analysis can be done by comparing our reddenings with other determinations. Table 9 summarises differences of analyzed tracers relative to our maps. The only significant difference exists for the comparison with Haschke et al. (2011) reddening maps, and is the result of adopted intrinsic color of the red clump. The 0.008 mag difference obtained with the Choi et al. 2019 maps based on the red clump E$(g-i)$ color, confirm that the red clump is affected by the reddening in a similar way as tracers used in this paper to obtain reddening zero-point. It is also worthwhile to mention that the average differences obtained for RR Lyrae stars and Cepheids are also small. We have to note, however, that distribution of reddening differences obtained for Cepheids and our map is highly asymmetrical, and should be analyzed further. We conclude this subsection by saying that we do not see systematic differences of reddening affecting analyzed in this subsection type of reddening tracers, regardless if they belong to early or late-type, neither to young, intermediate or old population.

\subsection{Reddening law}
The number of measurements suggests that the reddening law R$_V$ value for particular stars in our Galaxy or Magellanic Clouds can vary from 2.2 to 5.8. However, it is shown that for the diffuse interstellar medium for optical and near-infrared part of the spectrum, the mean value of the reddening law is R$_V$=3.1 (Cardelli et al. 1989; Pei 1992; Weingartner \& Drain 2001; Fitzpatrick \& Massa 2007;  Schlafly \& Finkbeiner 2011). In this subsection we investigate what would be the effect of a different reddening law R$_V$ value on our reddening maps. 
We recalculated our maps with the R$_V$=2.7 and R$_V$=4.5. Utilising Cardelli et al. (1989) reddening law parametrization, we obtain color excess transformation E$(V-I)$=1.227 $\times$ E$(B-V)$ for the R$_V$=2.7, and E$(V-I)$=1.635 $\times$ E$(B-V)$ for the R$_V$=4.5. 
This leads to the difference of the mean intrinsic $(V-I)_0$ color of the red clump. Despite the difference of the intrinsic red clump color, the differences of the average E$(B-V)$ reddening for the LMC are below 0.01 mag for R$_V$ 2.7, 3.1 and 4.5. Even for fields with reddening E$(B-V)$ $>$ 0.15 mag, the average difference in those fields is below 0.01 mag, and is within the systematic uncertainty of our maps.
The $(V-I)_0$ colors calculated separately for each method of reddening measurement, with the corresponding spread within each method, for R$_V$=2.7 and R$_V$=4.5 are reported in Table 10.

\section{Summary and Conclusions}
We have prepared reddening maps of the LMC and SMC. To obtain reddening map of both Magellanic Clouds, we measured the apparent color of the red clump in fields containing different types of reddening tracers. For each field with measured color we calculated the intrinsic, unreddened color of the red clump, by 
applying reddening measured with a tracer located in this field. We used three different types of reddening tracers: early-type eclipsing binaries, blue supergiants, and B-type stars with reddening measured with a Str\"{o}mgren  photometry. We obtained intrinsic color of the red clump $(V-I)_0 =  0.838 \pm 0.034 $ mag in the LMC, and $(V-I)_{0} =  0.815 \pm 0.034$ mag in the SMC. 
Those values are smaller than $(V-I)_0 =  0.92$  intrinsic color used by Haschke et al. (2011), and as a consequence result in higher values of the reddening with the offset E$(B-V)=0.061$ mag. We note that this difference transforms to 0.118 mag offset in the I-band if used directly to calibrate absolute brightness of any standard candle in the LMC. 
The minimal reddening obtained for the LMC is E$(B-V)=0.05$ mag and can be interpreted as the Galactic reddening toward LMC. The mean value of the reddening is $\langle$E$(B-V)\rangle=0.127$ and reaches maximum E$(B-V)=0.4$ mag in the fields of star forming regions close to 30 Doradus. Our maps clearly show clumpy and filamentous nature of the reddening in the LMC. We also compared our reddenings with numerous estimations based on different types of reddening tracers. Cepheids, RR Lyrae stars, early-type eclipsing binaries and other maps based on the red clump color on average show reddenings consistent with our map to within a few hundredths of magnitude. 
In the SMC, the minimal reddening is E$(B-V)=0.04$ mag and the maximum value is E$(B-V)=0.227$ mag. On average the reddening in the SMC 
is more homogeneous and smaller than in the LMC, with the average value $\langle$E$(B-V)\rangle=0.08$ mag.

\acknowledgments
The research leading to these results has received funding from the European Research Council (ERC) under the
European Union’s Horizon 2020 research and innovation program (grant agreement No 695099). We (W.G.,
G.P., D.G. and W.N.) also very gratefully acknowledge financial support for this work from the BASAL Centro de Astrofisica
y Tecnologias Afines (CATA) AFB-170002. We also acknowledge support from the IdP II 2015 0002 64 grant 
of the Polish Ministry of Science and Higher Education. M. T. acknowledges financial support from the Polish National Science
Center grant PRELUDIUM 2016/21/N/ST9/03310. M.G. gratefully acknowledges support from 
FONDECYT POSTDOCTORADO grant 3130361.

\newpage

\begin{figure}[ht!]
\plotone{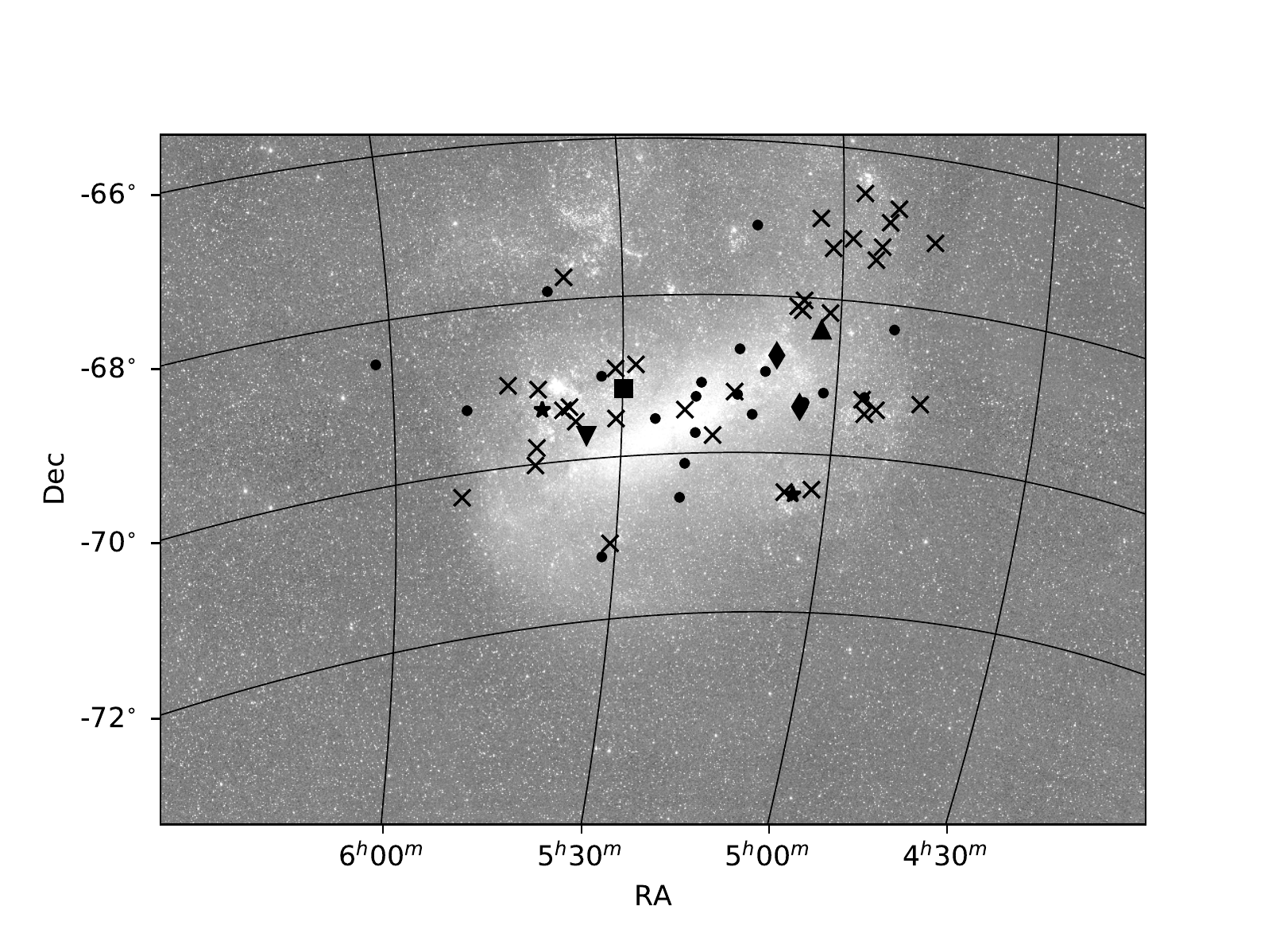}
\caption{Large Magellanic Cloud with marked positions of reddening tracers used in this paper to calculate unreddened, intrinsic $(V-I)_0$ color of the red clump. Cross - blue supergiants (Urbaneja et al. 2017);  dots - late type eclipsing binaries (Graczyk et al. 2018); square - Str\"{o}mgren photometry for B type stars (Larsen et al. 2000). 
Positions of reddening tracers used to compare with values obtained from our map are also marked: diamonds - clusters NGC 1850 (Lee 1995) and NGC 1835 (Walker 1993); early-type eclipsing binaries:  BLMC01 and BLMC02 (Taormina et al. 2019) - star symbols; HV 2274 (Groenewegen \& Salaris 2001) - triangles up; LMC-SC1-105 (Bonanos et al. 2011) - triangle down. The underlying image was obtained by The All Sky Automated Survey and originates from Udalski et al. (2008a)  }
\end{figure}


\begin{figure}[ht!]
\plotone{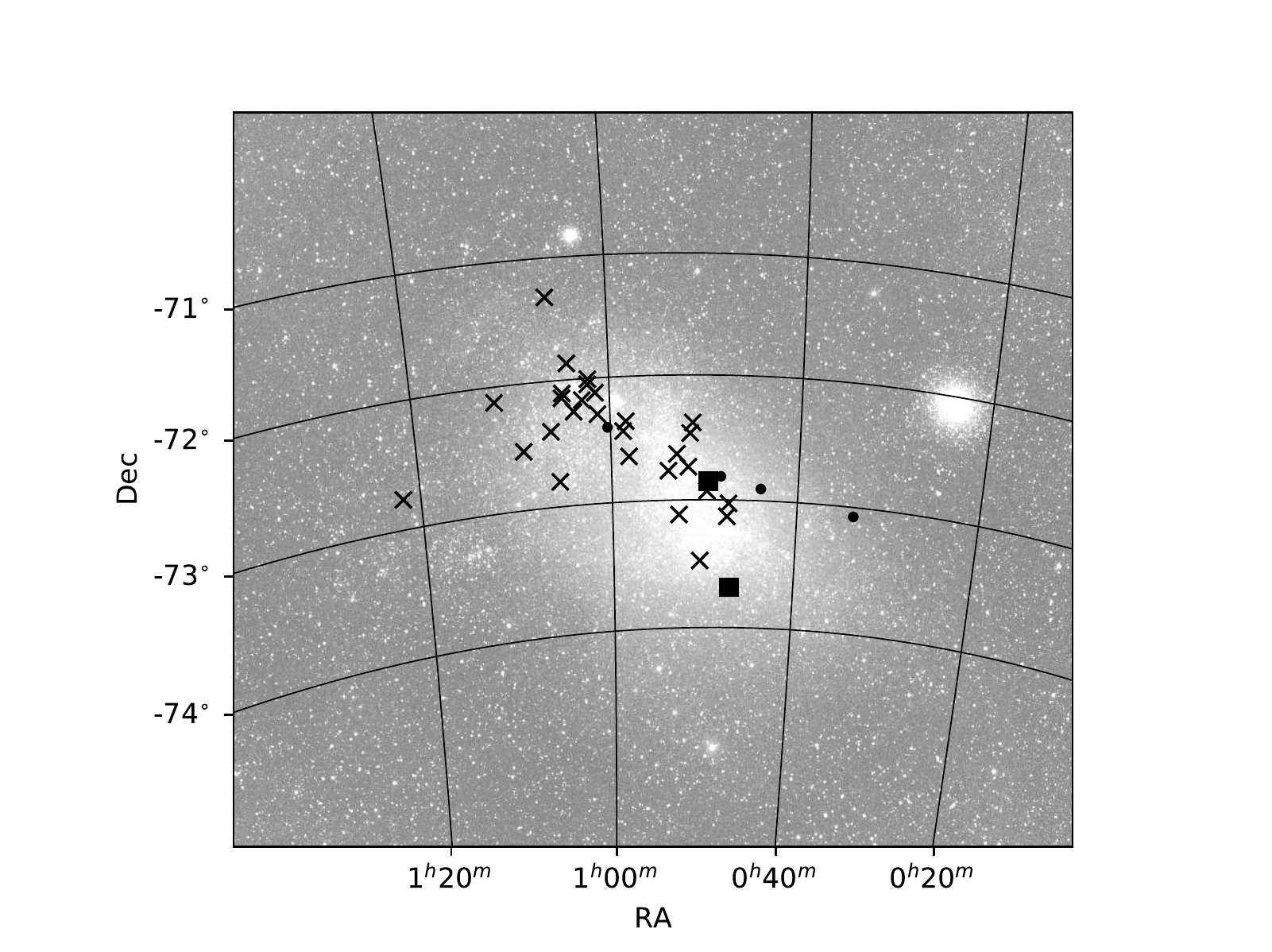}
\caption{Small Magellanic Cloud with marked positions of reddening measurements used in this paper to calculate unreddened, intrinsic $(V-I)_0$ color of the red clump. Cross - blue supergiants (Schiller 2010);  dots - late type eclipsing binaries (Graczyk et al. 2014);  square - Str\"{o}mgren photometry for B type stars (Larsen et al. 2000).
The underlying image was obtained by The All Sky Automated Survey and originate from Udalski et al. (2008b)}
\end{figure}


\begin{figure}[ht!]
\plotone{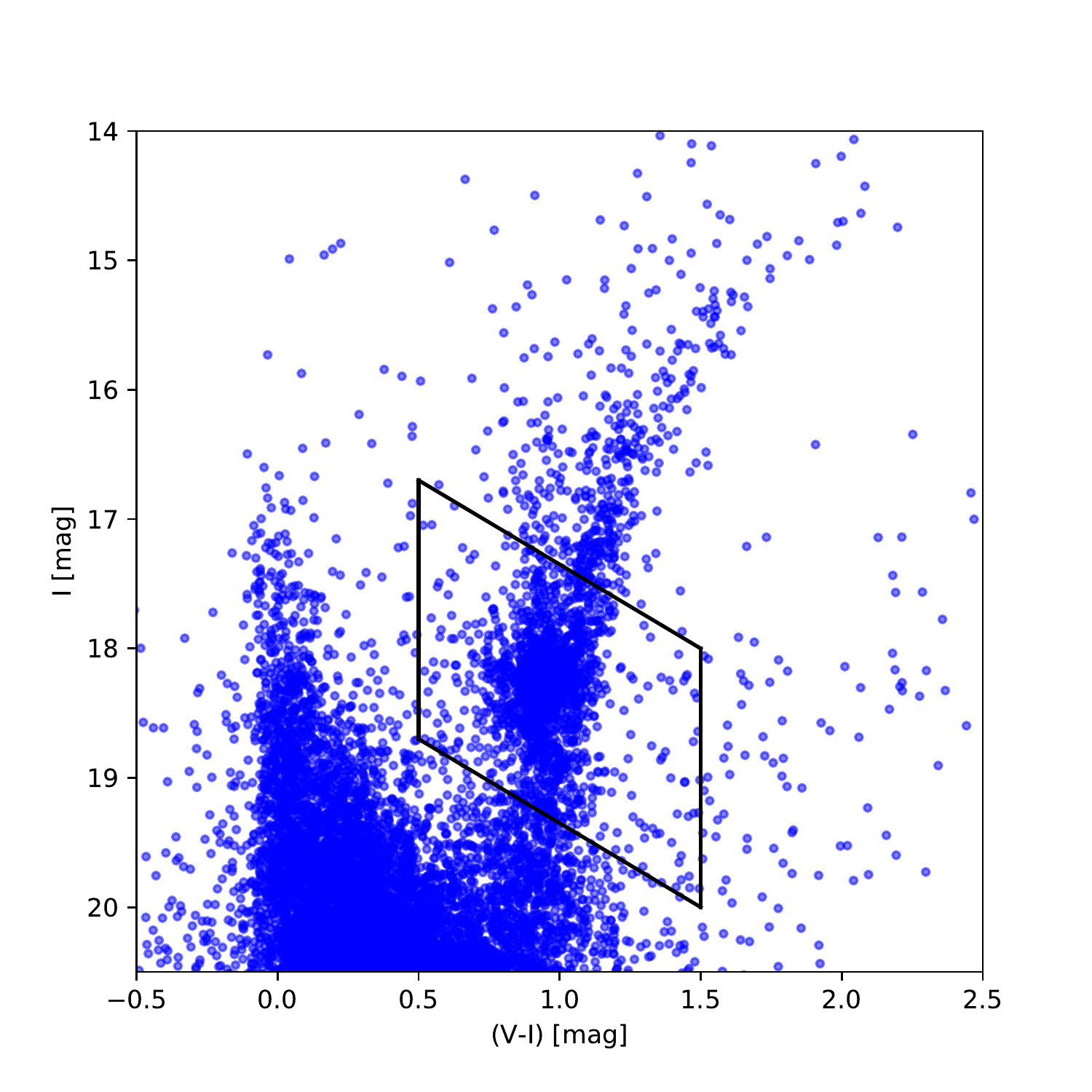}
\caption{Exemplary color-magnitude diagram of stars in the 5 $\times$ 5 arcmin field centered on system LMC-ECL-06575. The black box represents color and I- band magnitude red clump stars selection criteria.}
\end{figure}


\begin{figure}[ht!]
\plotone{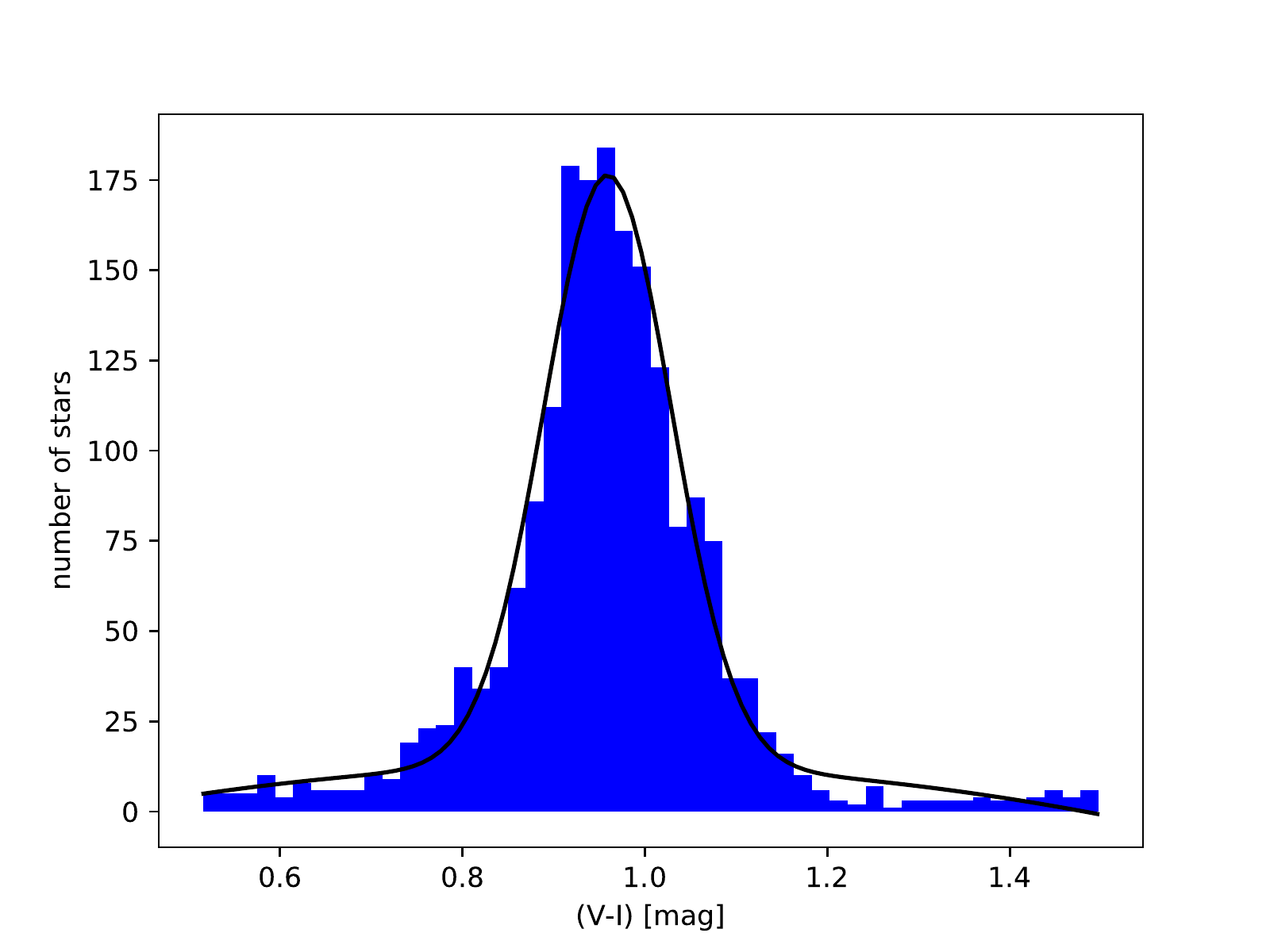}
\caption{Exemplary histogram of the colors of selected stars in the 5 $\times$ 5 arcmin field centered on system LMC-ECL-06575. Black line is the fit of the function (1), with the mean red clump color k$_{RC}$ = 0.960 $\pm$ 0.003 mag, red clump stars color spread $\sigma_{RC}$=0.072 mag and number of red clump stars N$_{RC}$=1434. }
\end{figure}


\begin{figure}[ht!]
\plotone{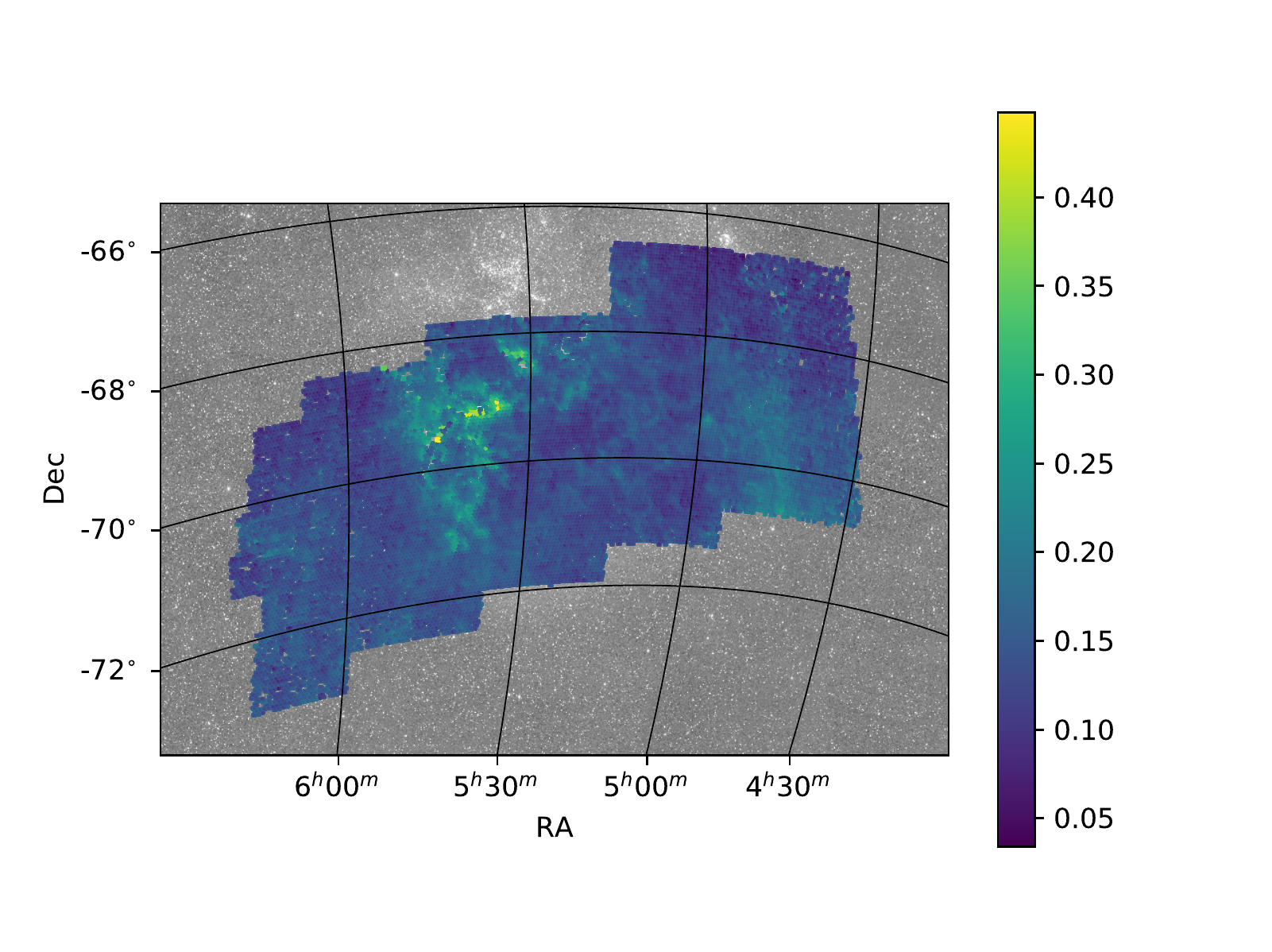}
\caption{Reddening map of the LMC. Measurements were performed in 3 $\times$ 3 arcmin fields. The color bar indicates reddening value E$(B-V)$ [mag]. }
\end{figure}


\begin{figure}[ht!]
\plotone{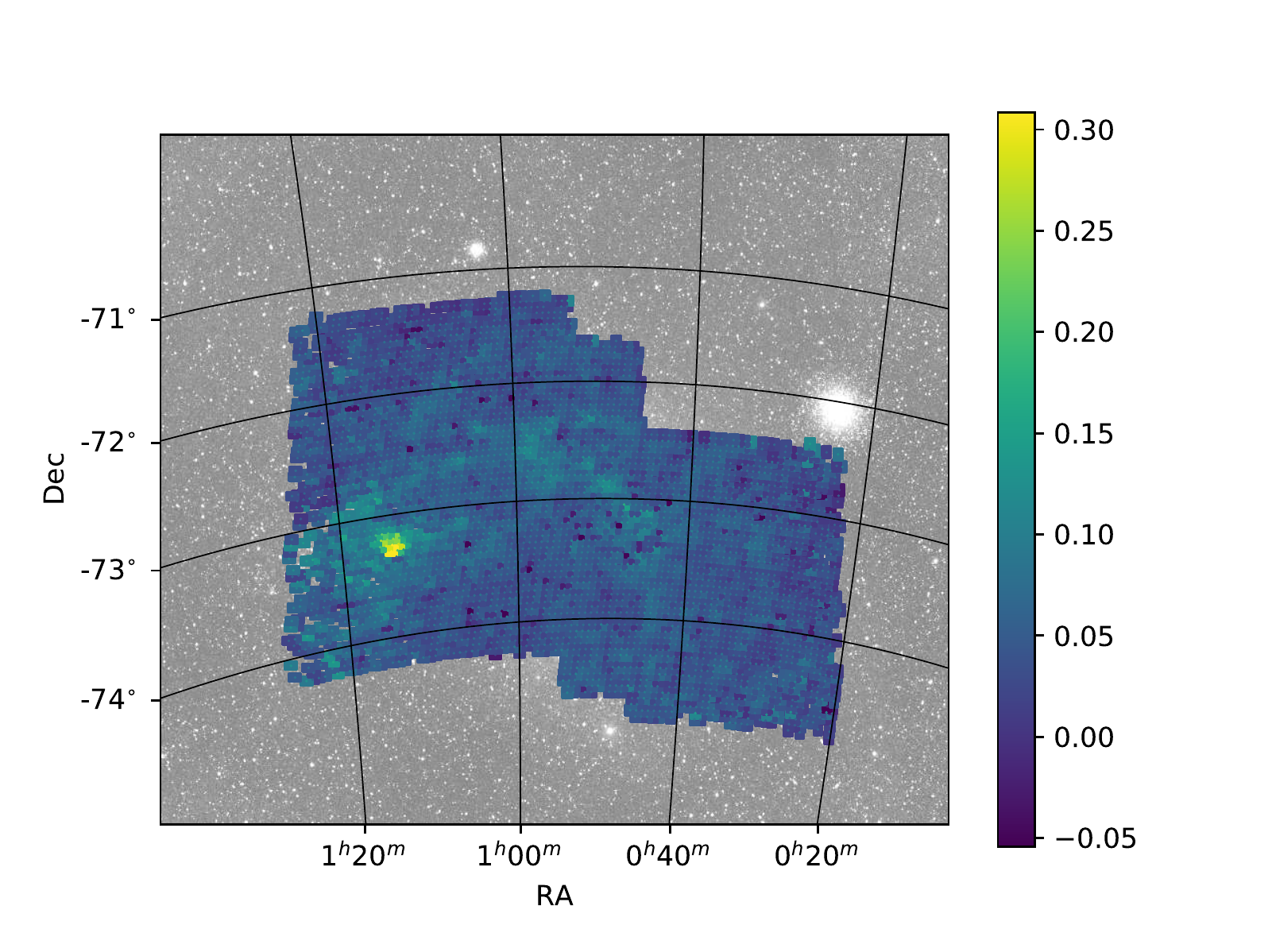}
\caption{Reddening map of the SMC. Measurements were performed in 3 $\times$ 3 arcmin fields. The color bar indicates reddening value E$(B-V)$ [mag]. }
\end{figure}


\begin{figure}[ht!]
\plotone{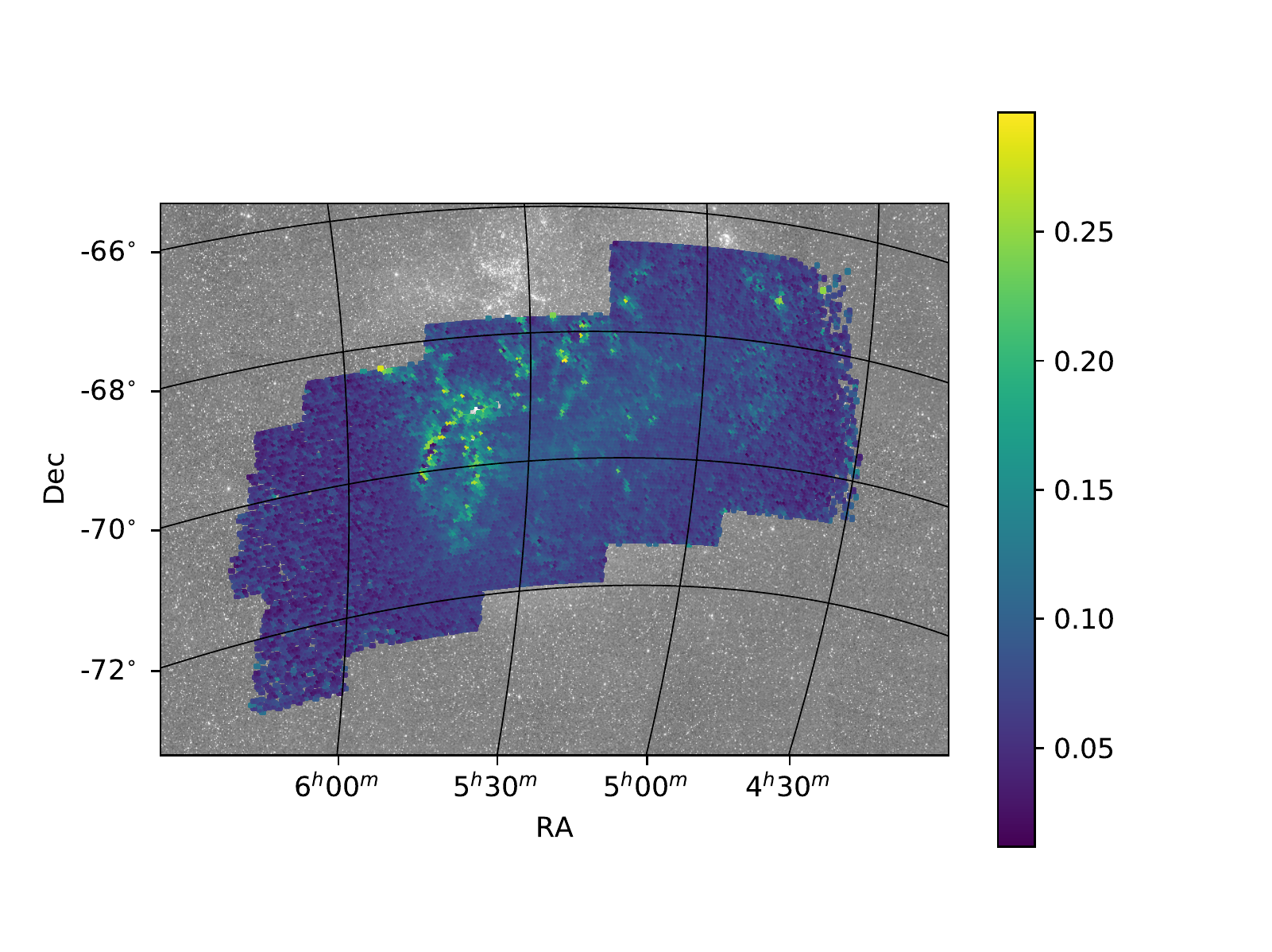}
\caption{Map of the red clump stars color spread ($\sigma_{RC}$) in the LMC. Measurements were performed for the same fields as for the reddening map.}
\end{figure}


\begin{figure}[ht!]
\plotone{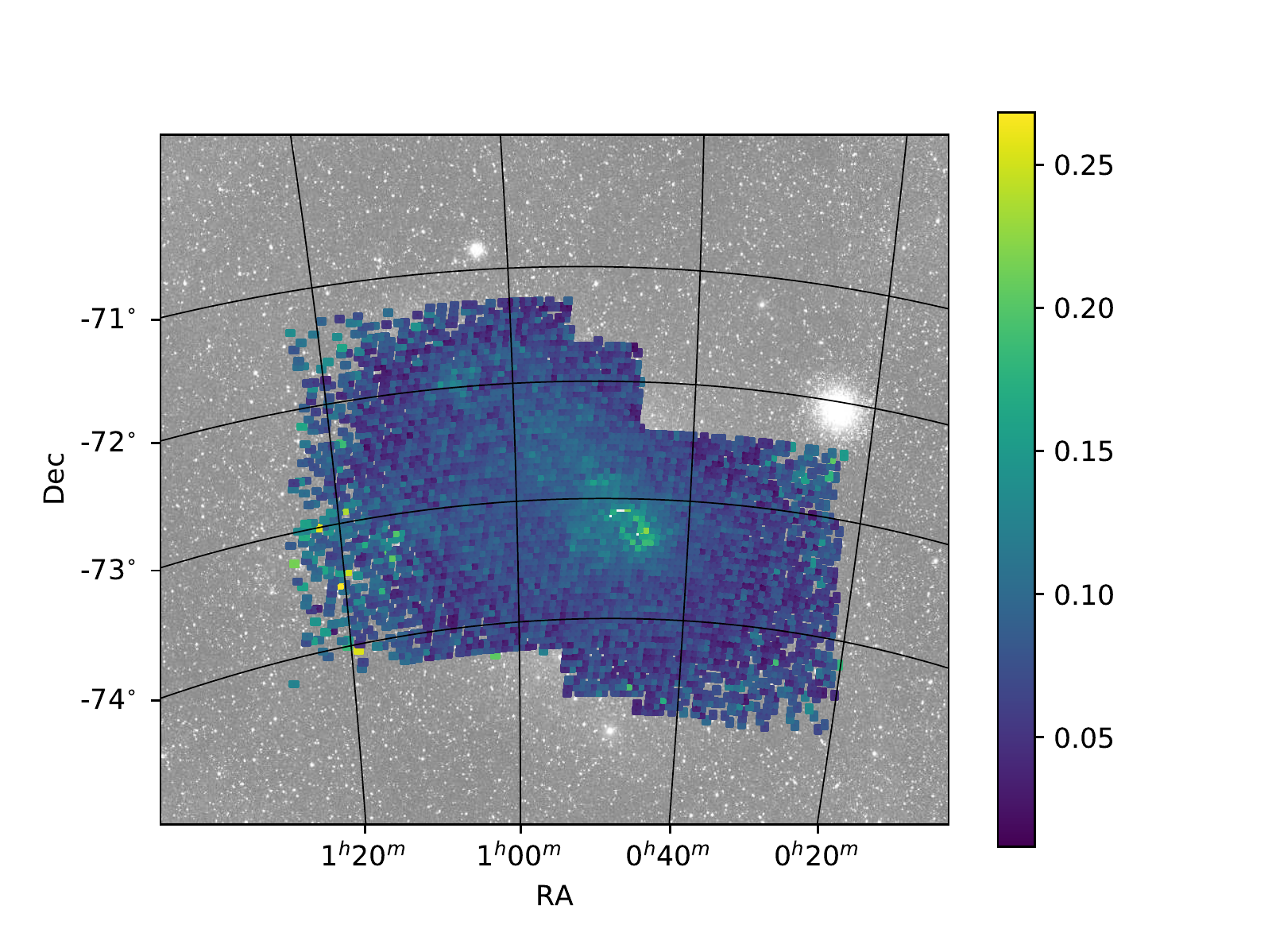}
\caption{Map of the red clump stars color spread ($\sigma_{RC}$) in the SMC. Measurements were performed for the same fields as for the reddening map.}
\end{figure}


\begin{figure}[ht!]
\plotone{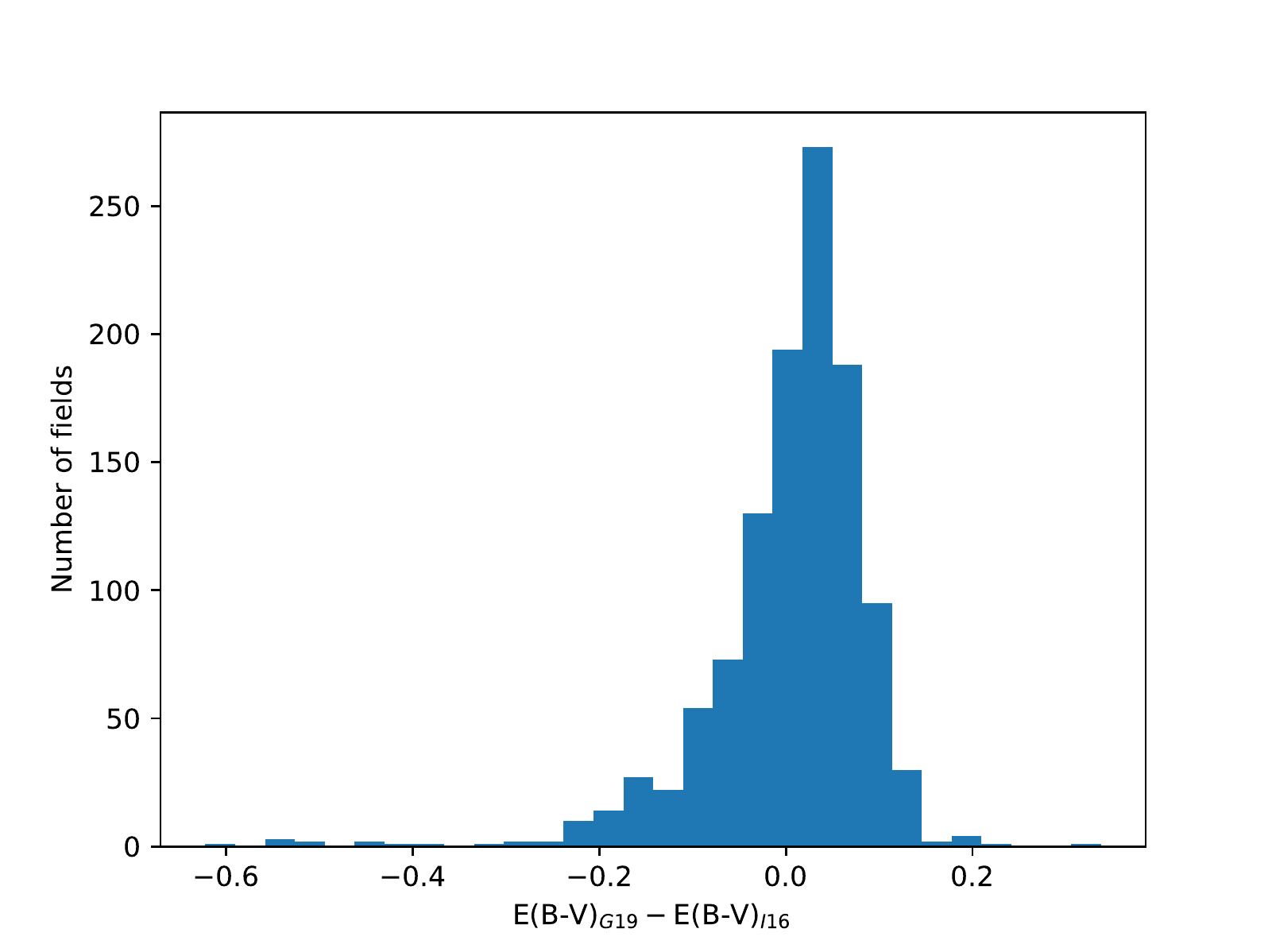}
\caption{Histogram of the difference of the reddening  E$(B-V)_{G19}$ obtained in this paper and reddening E$(B-V)_{I16}$ obtained by Inno et al. (2016) for 1133 LMC Cepheids pulsating in the fundamental mode. The mean value of the difference is 0.004 mag, with 0.087 mag spread.}
\end{figure}


\begin{figure}[ht!]
\plotone{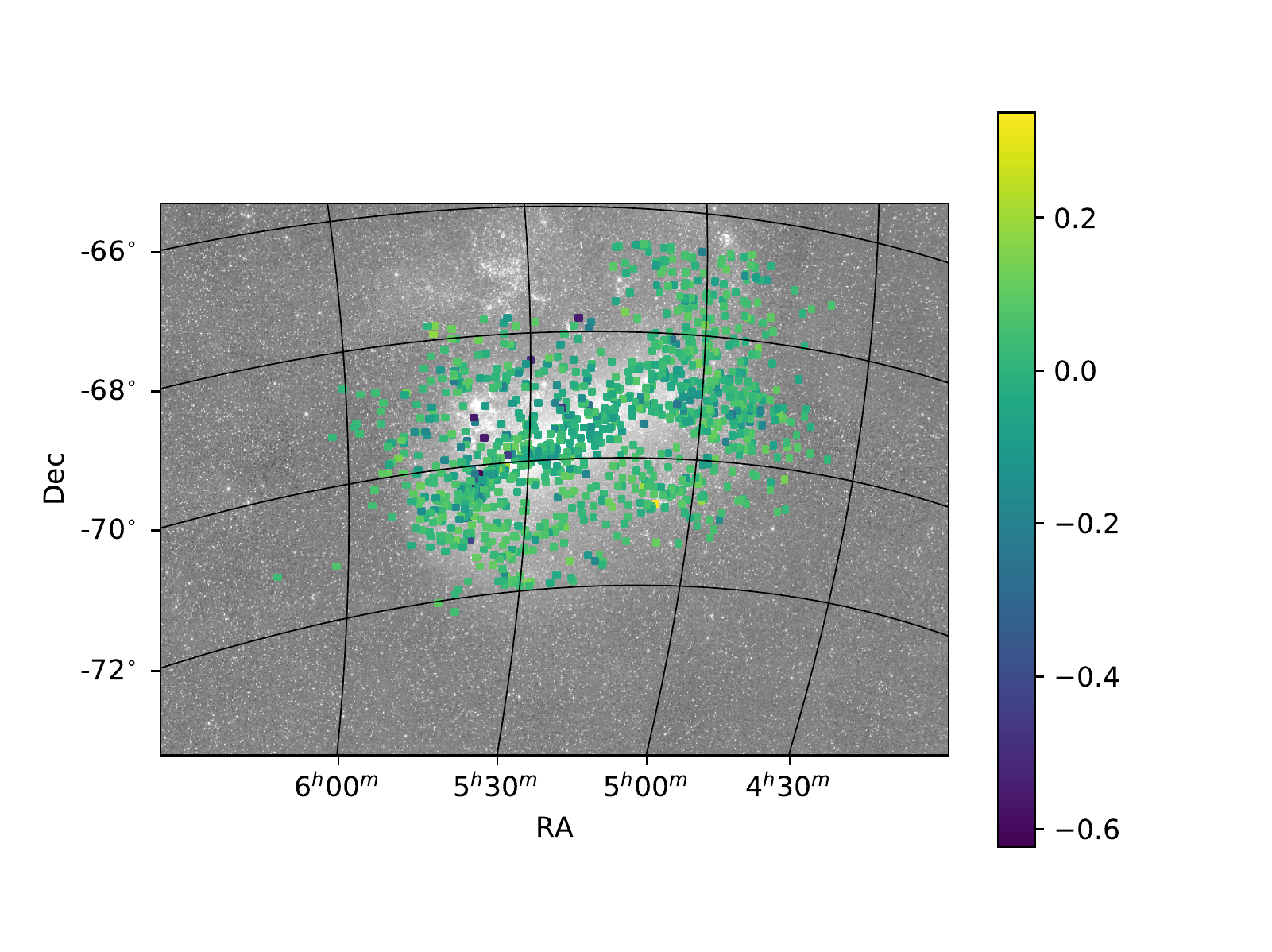}
\caption{Spatial distribution of the difference E$(B-V)_{G19}-$E$(B-V)_{I16}$ of the reddening obtained in this paper and reddening obtained by Inno et al. (2016) for the 1133 LMC Cepheids pulsating in fundamental mode.}
\end{figure}


\begin{figure}[ht!]
\plotone{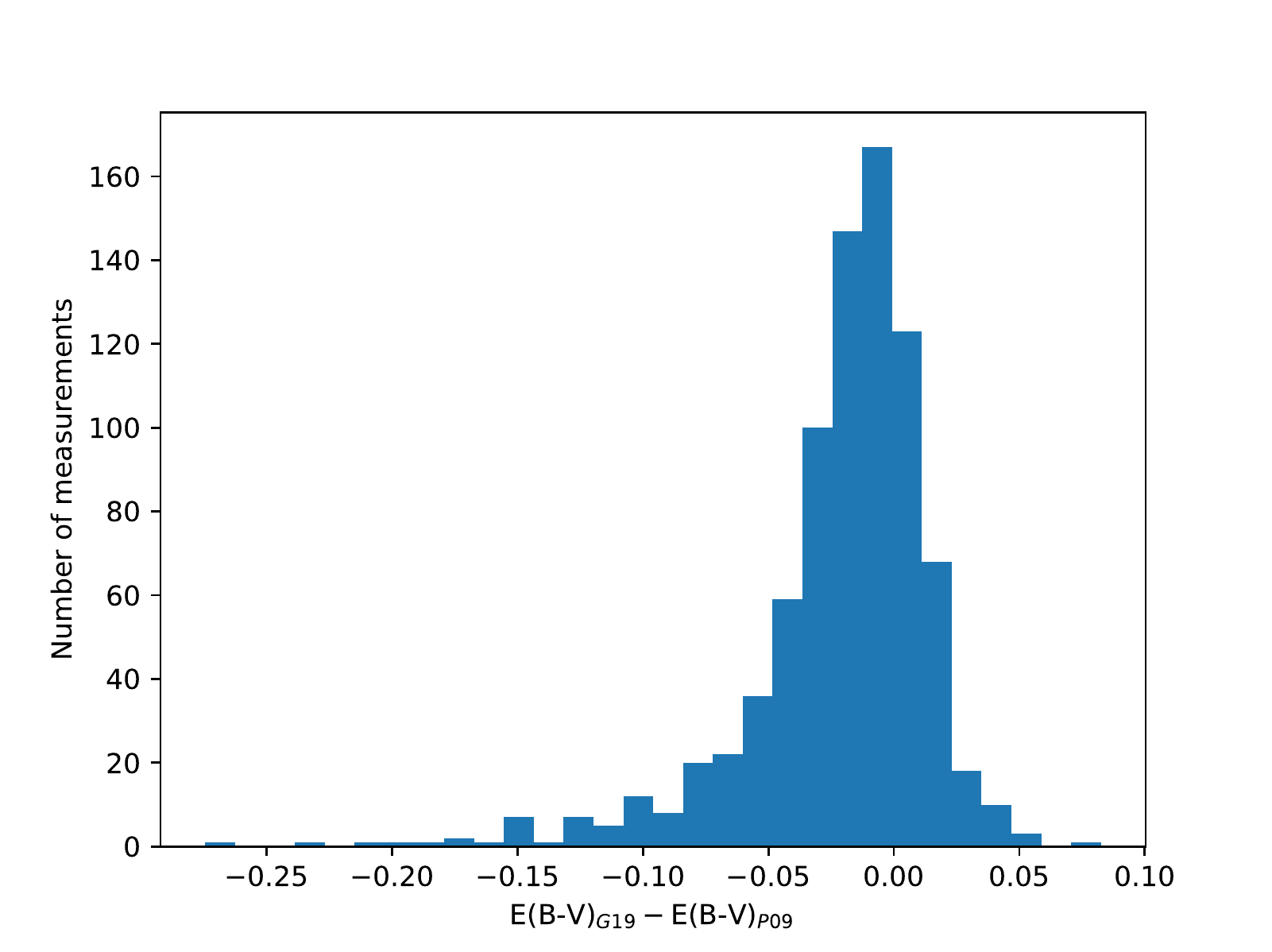}
\caption{Histogram of the difference of the reddening  E$(B-V)_{G19}$ obtained in this paper and reddening E$(B-V)_{P09}$ obtained by Pejcha \& Stanek (2009) for the RR Lyrae stars in the LMC. The mean value of the difference is -0.024 mag, with 0.02 mag spread.}
\end{figure}


\begin{figure}[ht!]
\plotone{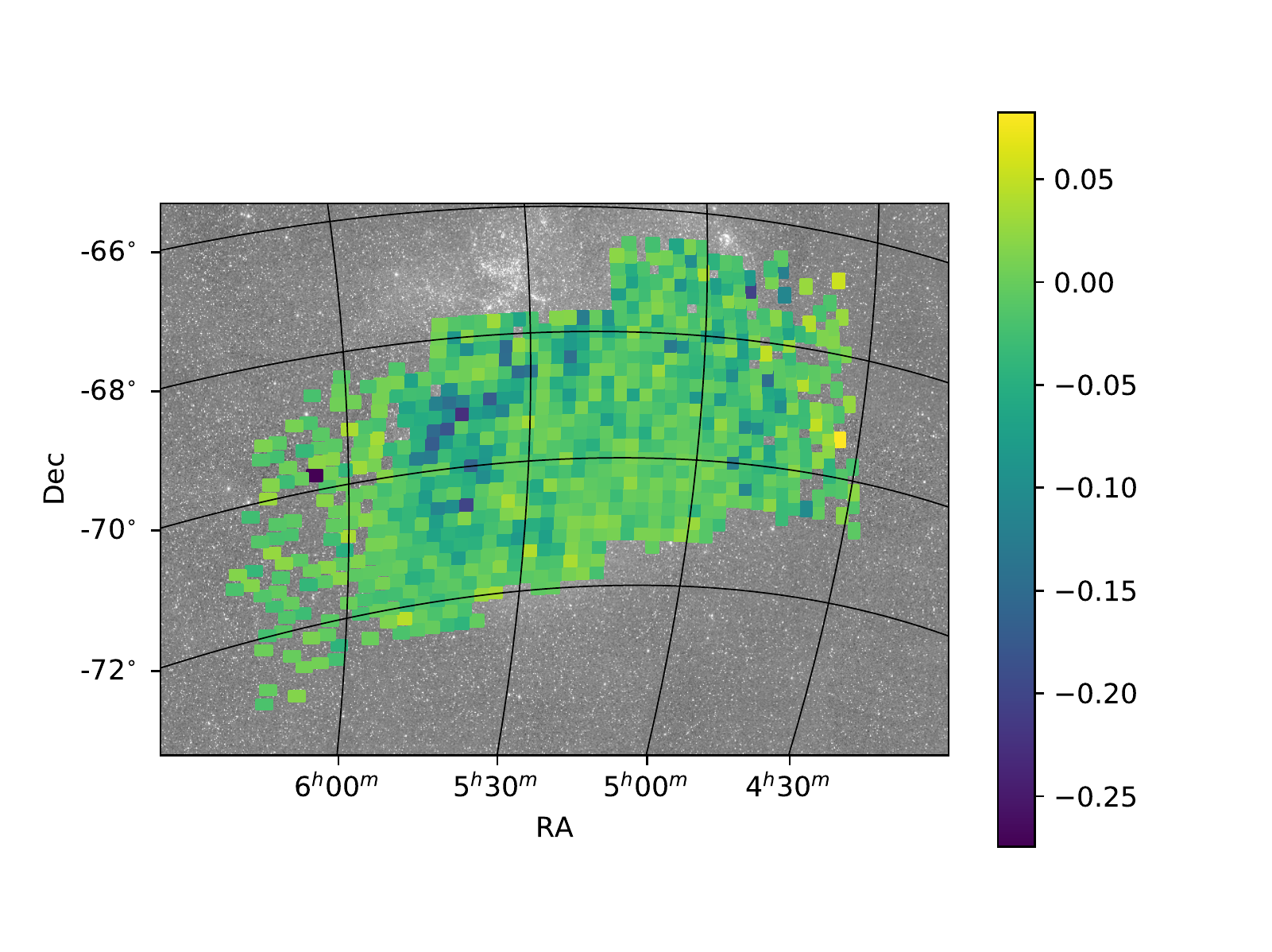}
\caption{Spatial distribution of the difference E$(B-V)_{G19}-$E$(B-V)_{P09}$ of the reddening obtained in this paper and reddening obtained by Pejcha \& Stanek (2009) for the RR Lyrae stars in the LMC.}
\end{figure}


\begin{figure}[ht!]
\plotone{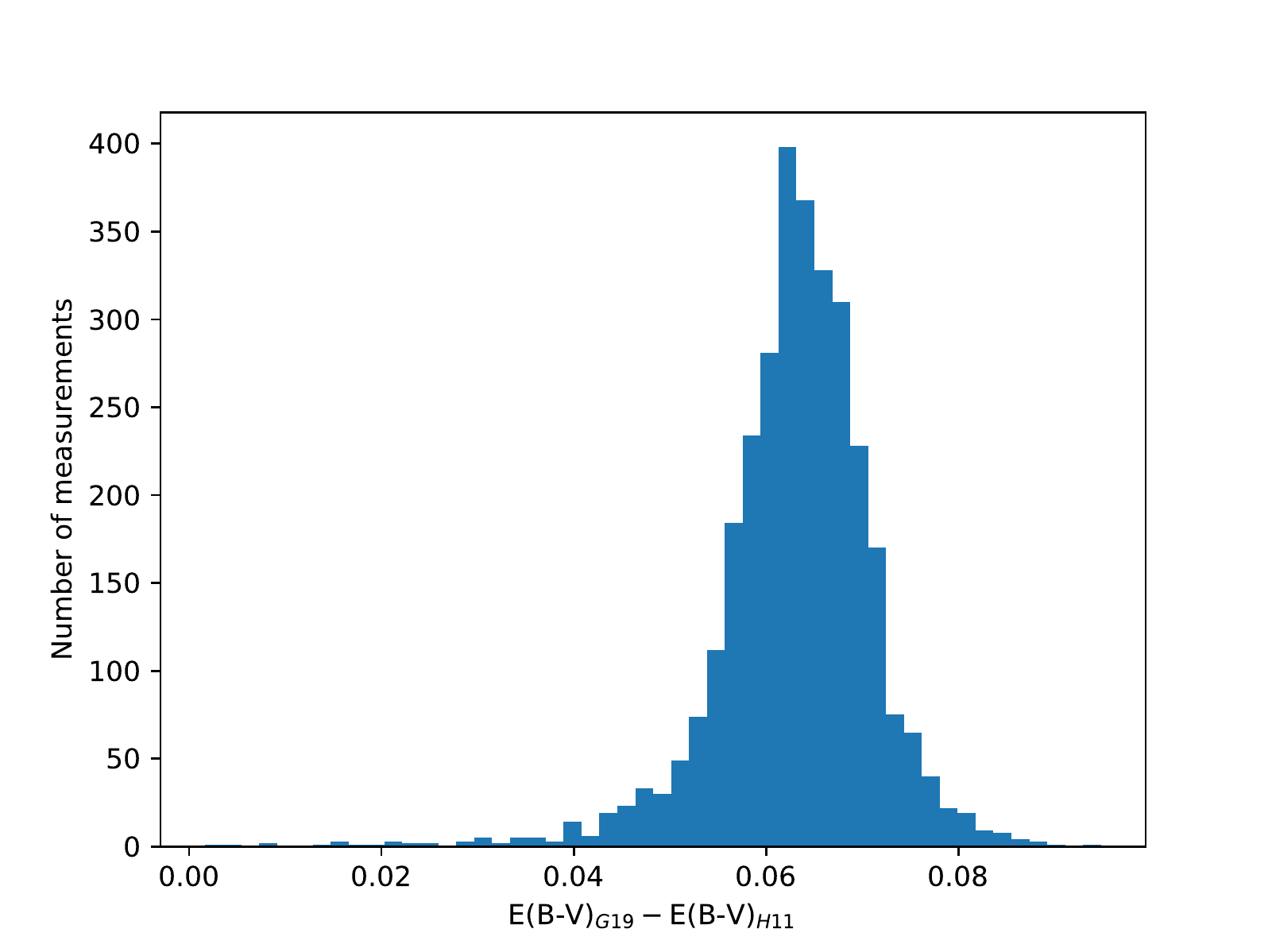}
\caption{Histogram of the difference of the reddening E$(B-V)_{G19}$ obtained in this paper and reddening E$(B-V)_{H11}$ obtained by Haschke et al. (2011) for the LMC. The 0.061 mag offset is clearly visible, and corresponds to the difference of the $(V-I)_0$ intrinsic red clump color adopted by Haschke et al. (2011) and red clump intrinsic color adopted in this paper. The 0.011 mag spread is due to different red clump star selection methods and statistical uncertainty of the measurement. }
\end{figure}


\begin{figure}[ht!]
\plotone{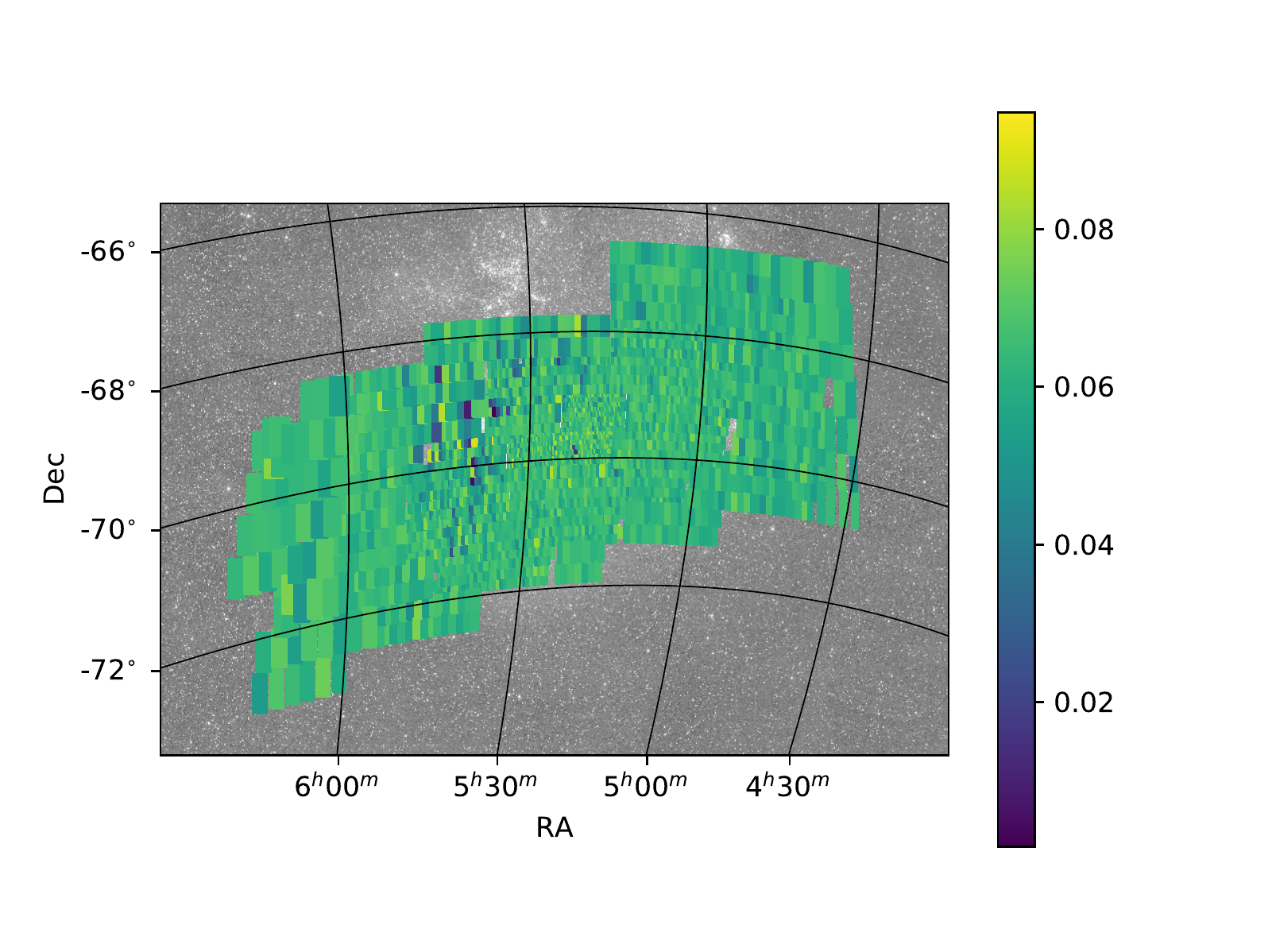}
\caption{
Spatial distribution of the difference E$(B-V)_{G19}-$E$(B-V)_{H11}$ of the reddening obtained in this paper, and reddening obtained by Haschke et al. (2011) for the LMC.} 
\end{figure}


\begin{figure}[ht!]
\plotone{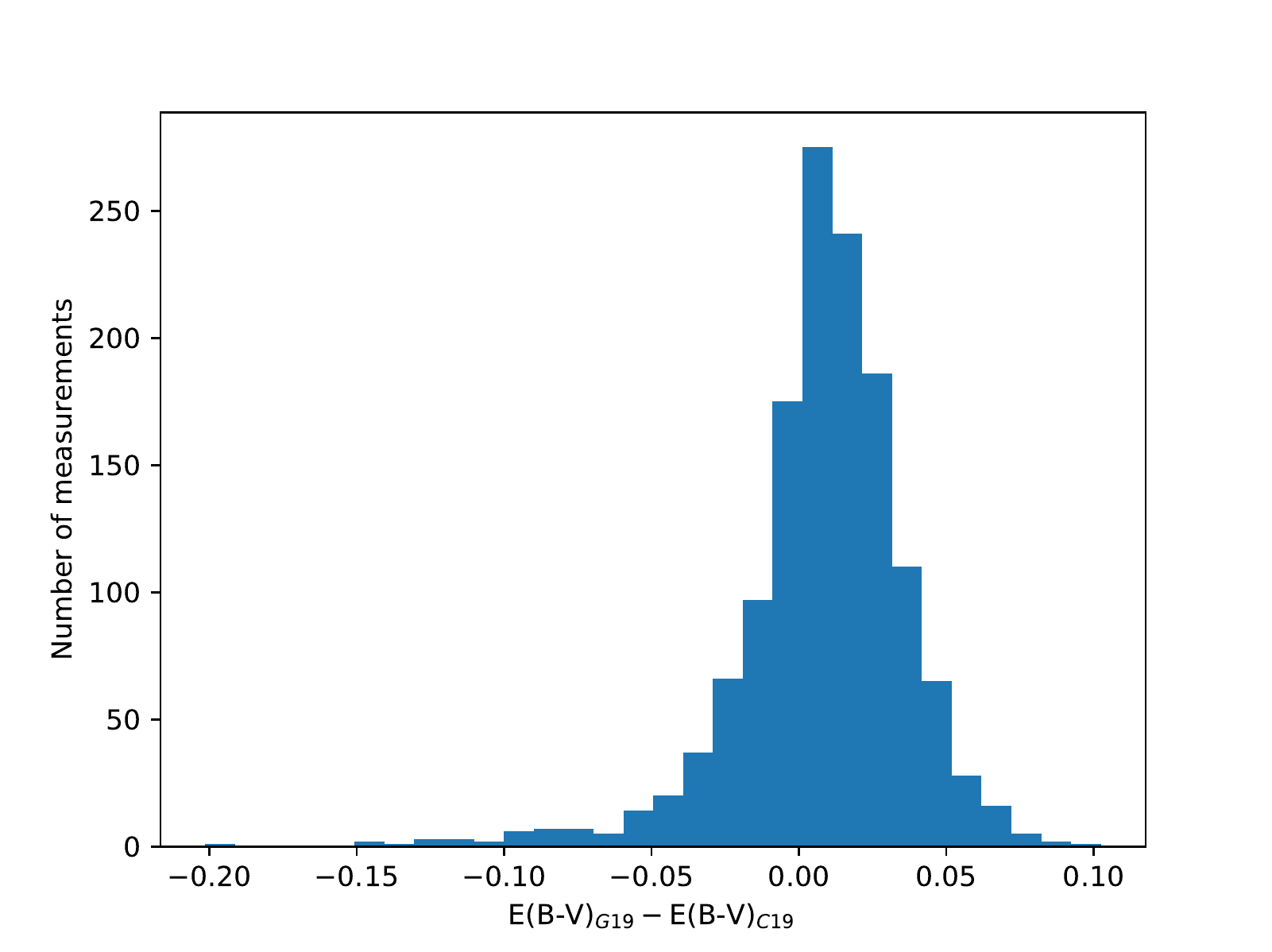}
\caption{Histogram of the difference of the reddening E$(B-V)_{G19}$ obtained in this paper and reddening E$(B-V)_{C19}$ obtained by Choi et al. (2019) for the LMC.}
\end{figure}


\begin{figure}[ht!]
\plotone{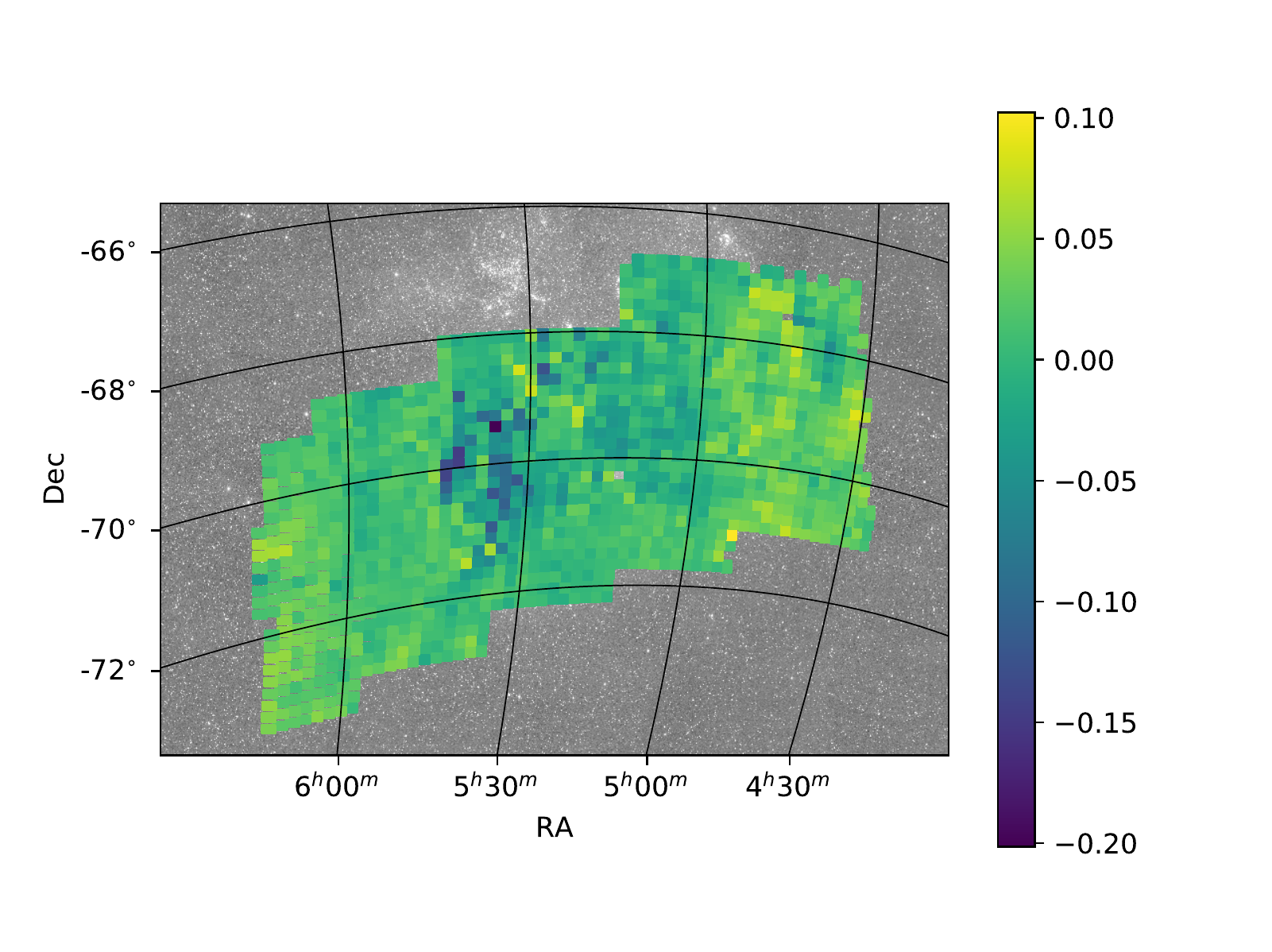}
\caption{
Spatial distribution of the difference E$(B-V)_{G19}-$E$(B-V)_{C19}$ of the reddening obtained in this paper, and reddening obtained by Choi et al. (2019) for the LMC.} 
\end{figure}


\begin{figure}[ht!]
\plotone{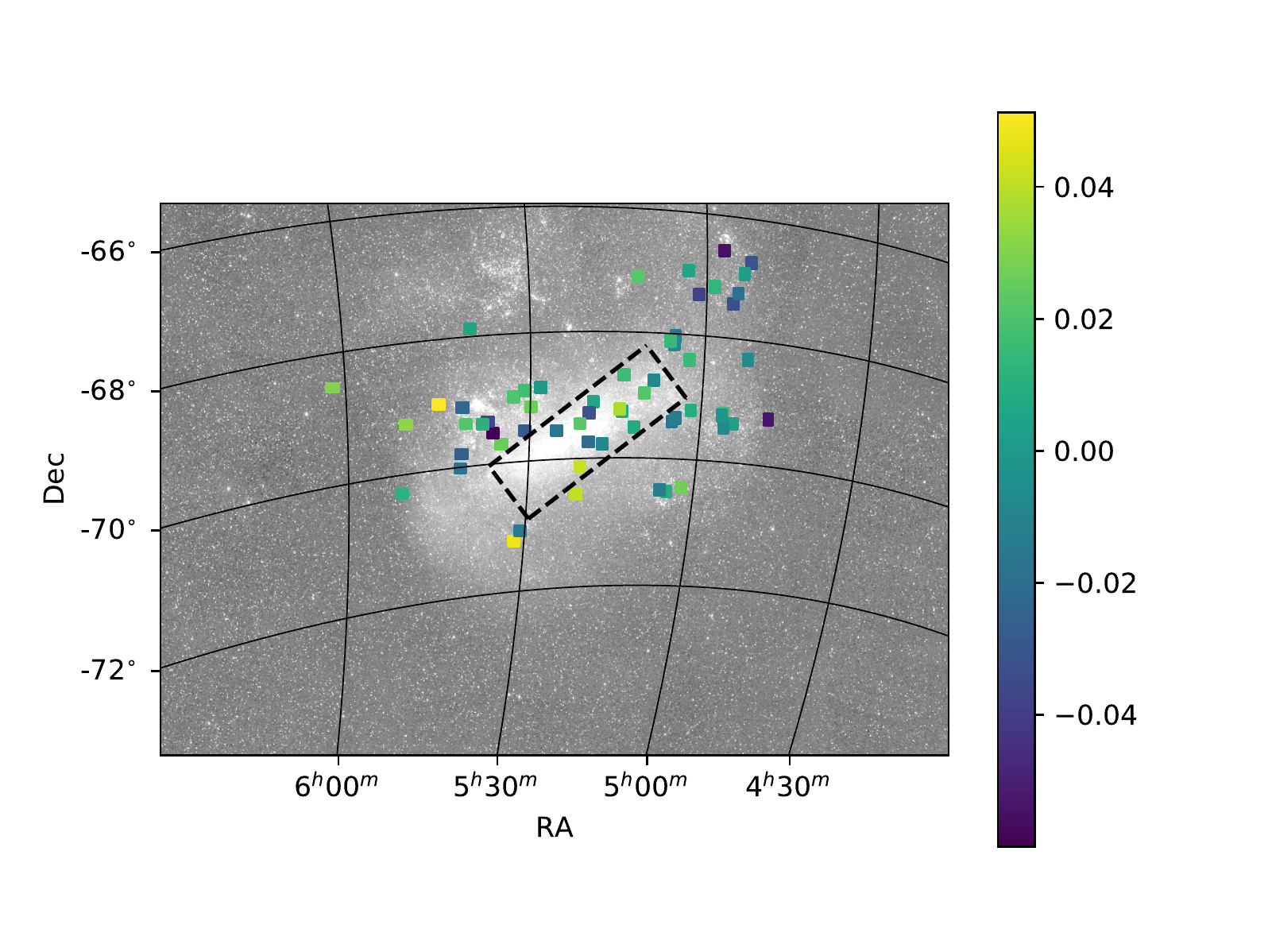}
\caption{
Spatial distribution of reddening tracers used to obtain $(V-I)_0$ intrinsic color of the red clump, plotted over the LMC. Six additional fields containing reddening tracers used to 
compare reddening obtained from our maps are also marked. Color bar indicates the difference of the reddening obtained from red clump color measured in each field, and 
value of the reddening for individual tracer E$(B-V)_{G19}-$E$(B-V)_{T}$. The dashed box encloses bar of the LMC. The average difference E$(B-V)_{G19}-$E$(B-V)_{T}$ for tracers inside the box is shifted by 0.005 mag in comparison to average value of all tracers. The uncertainty of this value is 0.006 mag, therefore, we conclude that there is no systematic difference of the $(V-I)_0$ red clump color in the bar of the LMC, and regions sampled with the remaining reddening tracers. 
} 
\end{figure}

\newpage

\begin{deluxetable}{cccccc}
\tablecaption{Reddening values obtained from the analysis of late-type eclipsing binaries in the LMC and SMC (Graczyk et al. 2018; Graczyk et al. 2014). In the last column we report measured $(V-I)$ color of the red clump.}
\tablehead{
\colhead{ID} & \colhead{R.A.} & \colhead{Decl.} & \colhead{E$(B-V)$} & \colhead{E$(B-V)$}  & \colhead{$(V-I)_{RC}$} \\
\colhead{} & \colhead{} & \colhead{} & \colhead{Na I} & \colhead{Atm.}  & \colhead{}
}
\startdata
\multicolumn{6}{c}{LMC} \\
\hline
LMC-ECL-01866  &  04:52:15.28  &  -68:19:10.3 & 0.134  &  0.094    & 1.006 $\pm$ 0.004 \\
LMC-ECL-03160  &  04:55:51.48  &  -69:13:48.0 & 0.109  &  0.151    & 1.005 $\pm$ 0.003 \\
LMC-ECL-05430  &  05:01:51.74  &  -69:12:48.8 & 0.108  &  0.120    & 0.992 $\pm$ 0.003 \\
LMC-ECL-06575  &  05:04:32.87  &  -69:20:51.0 & 0.105  &  0.134    & 0.960 $\pm$ 0.003 \\
LMC-ECL-09114  &  05:10:19.64   &  -68:58:12.2 & 0.091  &  0.114   & 0.986 $\pm$ 0.004 \\
LMC-ECL-09660  &  05:11:49.45  &  -67:05:45.2 & 0.093  &  0.125    & 0.989 $\pm$ 0.005 \\
LMC-ECL-09678  &  05:11:51.76  &  -69:31:01.1 & 0.143  &  0.117    & 1.036 $\pm$ 0.005 \\
LMC-ECL-10567  &  05:14:01.89  &  -68:41:18.2 & 0.097  &  0.119    & 0.987 $\pm$ 0.002 \\
LMC-SC9-230659 &  05:14:06.04  &  -69:15:56.9 & 0.126  &  0.183    & 1.013 $\pm$ 0.005 \\
LMC-ECL-12669  &  05:19:12.80  &  -69:06:44.4 & 0.122  &  0.074    & 1.006 $\pm$ 0.003 \\
LMC-ECL-12875  &  05:19:45.39  &  -69:44:38.5 & 0.122  &  0.210    & 0.971 $\pm$ 0.002 \\
LMC-ECL-12933  &  05:19:53.69  &  -69:17:20.4 & 0.124  &  0.186    & 0.959 $\pm$ 0.003 \\
LMC-ECL-13360  &  05:20:59.46  &  -70:07:35.2 & 0.112  &  0.159    & 1.041 $\pm$ 0.003 \\
LMC-ECL-13529  &  05:21:23.34  &  -70:33:00.0 & 0.100  &  0.132    & 1.024 $\pm$ 0.003 \\
LMC-ECL-15260  &  05:25:25.66  &  -69:33:04.5 & 0.106  &  0.125    & 0.956 $\pm$ 0.002 \\
LMC-ECL-18365  &  05:31:49.56  &  -71:13:28.3 & 0.101  &  0.149    & 1.035 $\pm$ 0.002 \\
LMC-ECL-18836  &  05:32:53.06  &  -68:59:12.3 & 0.167  &  0.189    & 1.085 $\pm$ 0.007 \\
LMC-ECL-21873  &  05:39:51.19  &  -67:53:00.5 & 0.115  &  0.127    & 0.998 $\pm$ 0.004 \\
LMC-ECL-24887  &  05:50:39.02  &  -69:14:20.7 & 0.139  &  0.202    & 1.064 $\pm$ 0.003 \\
LMC-ECL-25658  &  06:01:58.77  &  -68:30:55.1 & 0.065  &  0.106    & 0.964 $\pm$ 0.003 \\
\hline
\multicolumn{6}{c}{SMC} \\
\hline
SMC101.8-14077 & 00:48:22.70 & -72:48:48.6 & 0.078 & 0.046 & 0.952 $\pm$ 0.008 \\ 
SMC108.1-14904 & 01:00:18.10 & -72:24:07.1 &  0.085 & 0.107 & 0.938 $\pm$ 0.006 \\ 
SMC126.1-210 & 00:44:02.68 & -72:54:22.5 & 0.084 & 0.097   & 0.908 $\pm$ 0.005 \\ 
SMC130.5-4296 & 00:33:47.90 & -73:04:28.0 &  0.063 & 0.101 & 0.936 $\pm$ 0.005 \\ 
\enddata
\end{deluxetable}


\begin{deluxetable}{ccc|cc}
\tablecaption{E$(V-I)$ color excess values in the LMC obtained for blue supergiants (Urbaneja et al. 2017). The E$(V-I)$ values were obtained with Cardelli et al. (1989) reddening law parametrization with the E(4405-5495) color excess and R$_{5495}$ reddening law values from Urbaneja et al. (2017). In the last column we report measured $(V-I)$ color of the red clump.}
\tablehead{
\colhead{ID} & \colhead{R.A.} & \colhead{Decl.} &    \colhead{E$(V-I)$}    & \colhead{$(V-I)_{RC}$}

}
\startdata
Sk-66-50 &  05:03:08.82368 &  -66:57:34.8883 & 0.127  & 0.971 $\pm$ 0.005 \\
Sk-67-19 &  04:55:21.60684 &  -67:26:11.2579 & 0.179  & 0.973 $\pm$ 0.005 \\
Sk-69-2A &  04:47:33.22400 &  -69:14:32.8700 & 0.306  & 1.073 $\pm$ 0.008 \\
Sk-69-24 &  04:53:59.51900 &  -69:22:42.5700 & 0.203  & 1.045 $\pm$ 0.007 \\
Sk-69-39A &  04:55:40.44700 &  -69:26:40.9800 & 0.205 & 1.033 $\pm$ 0.004 \\
Sk-69-82 &  05:14:31.56600 &  -69:13:53.3900 & 0.117  & 1.004 $\pm$ 0.004 \\
Sk-69-113 &  05:21:22.39900 &  -69:27:08.0700 & 0.094 & 0.962 $\pm$ 0.002 \\
Sk-69-170 &  05:30:50.07611 &  -69:31:29.3838 & 0.197 & 0.997 $\pm$ 0.002 \\
Sk-69-299 &  05:45:16.61696 &  -68:59:51.9691 & 0.228 & 1.134 $\pm$ 0.003 \\
Sk-70-45 &  05:02:17.80400 &  -70:26:56.7400 & 0.101  & 0.975 $\pm$ 0.003 \\
Sk-66-1 &  04:52:19.05300 &  -66:43:53.2300 & 0.180   & 0.976 $\pm$ 0.006 \\
Sk-67-14 &  04:54:31.89070 &  -67:15:24.6640 & 0.142  & 0.956 $\pm$ 0.008 \\
Sk-67-36 &  05:01:22.59000 &  -67:20:10.0200 & 0.150  & 0.937 $\pm$ 0.003 \\
Sk-67-228 &  05:37:41.01200 &  -67:43:16.5600 & 0.176 & 0.980 $\pm$ 0.002 \\
Sk-68-40 &  05:05:15.20300 &  -68:02:14.1500 & 0.163  & 0.981 $\pm$ 0.003 \\
Sk-68-92 &  05:28:16.15700 &  -68:51:45.5800 & 0.156  & 0.994 $\pm$ 0.002 \\
Sk-69-43 &  04:56:10.45700 &  -69:15:38.2100 & 0.162  & 0.997 $\pm$ 0.002 \\
Sk-66-5 &  04:53:30.02900 &  -66:55:28.2400 & 0.134   & 0.974 $\pm$ 0.010 \\
Sk-66-35 &  04:57:04.44000 &  -66:34:38.4500 & 0.169  & 0.934 $\pm$ 0.005 \\
Sk-67-28 &  04:58:39.23000 &  -67:11:18.7000 & 0.101  & 0.957 $\pm$ 0.002 \\
Sk-68-41 &  05:05:27.13100 &  -68:10:02.6100 & 0.134  & 0.960 $\pm$ 0.009 \\
Sk-68-45 &  05:06:07.29500 &  -68:07:06.1100 & 0.104  & 0.962 $\pm$ 0.014 \\
Sk-68-111 &  05:31:00.84200 &  -68:53:57.1700 & 0.153 & 1.014 $\pm$ 0.003 \\
Sk-69-89 &  05:17:17.57910 &  -69:46:44.1650 & 0.161  & 0.988 $\pm$ 0.002 \\
Sk-69-214 &  05:36:16.43600 &  -69:31:27.0900 & 0.290 & 1.049 $\pm$ 0.003 \\
Sk-69-228 &  05:37:09.21800 &  -69:20:19.5400 & 0.305 & 1.096 $\pm$ 0.004 \\
Sk-69-237 &  05:38:01.31260 &  -69:22:14.0790 & 0.238 & 1.091 $\pm$ 0.007 \\
Sk-69-270 &  05:41:20.40800 &  -69:05:07.3600 & 0.332 & 1.140 $\pm$ 0.005 \\
Sk-69-274 &  05:41:27.67900 &  -69:48:03.7000 & 0.245 & 1.047 $\pm$ 0.007 \\
Sk-70-78 &  05:06:16.03500 &  -70:29:35.7600 & 0.152  & 0.973 $\pm$ 0.003 \\
Sk-70-111 &  05:41:36.79000 &  -70:00:52.6500 & 0.215 & 1.032 $\pm$ 0.004 \\
Sk-70-120 &  05:51:20.78027 &  -70:17:09.3281 & 0.152 & 1.005 $\pm$ 0.002 \\
Sk-71-42 &  05:30:47.77615 &  -71:04:02.2976 & 0.198  & 1.017 $\pm$ 0.005 \\
\enddata
\end{deluxetable}


\begin{deluxetable}{ccc|cc}
\tablecaption{Reddening values in the SMC obtained for blue supergiants (Schiller 2010). In the last column we report measured $(V-I)$ color of the red clump.}
\tablehead{
\colhead{ID} & \colhead{R.A.} & \colhead{Decl.} &    \colhead{E$(B-V)$}  & \colhead{$(V-I)_{RC}$}
}
\startdata
AV20 & 00:47:29.21578 &  -73:01:37.3645 &  0.250  & 0.970 $\pm$ 0.011 \\ 
AV22 & 00:47:38.74027 &  -73:07:48.8527 &  0.100  & 0.910 $\pm$ 0.095 \\ 
AV56 & 00:49:51.27370 &  -72:55:45.2540 &  0.130  & 0.986 $\pm$ 0.005 \\ 
AV76 & 00:50:31.58196 &  -73:28:42.5531 &  0.100  & 0.933 $\pm$ 0.004 \\ 
AV98 & 00:51:24.59727 &  -72:22:58.5985 &  0.060  & 0.921 $\pm$ 0.006 \\ 
AV105 & 00:51:41.23958 &  -72:28:06.6951 &  0.060 & 0.917 $\pm$ 0.004 \\
AV110 & 00:51:51.98495 &  -72:44:13.5272 &  0.080 & 0.966 $\pm$ 0.007 \\
AV136 & 00:52:51.23667 &  -73:06:53.6430 &  0.070 & 0.930 $\pm$ 0.006 \\
SK56 & 00:53:04.88825 &  -72:38:00.1420 &  0.110  & 0.931 $\pm$ 0.004 \\ 
AV151 & 00:53:59.37458 &  -72:45:59.6595 &  0.140 & 0.948 $\pm$ 0.004 \\
AV200 & 00:58:07.90615 &  -72:38:30.4920 &  0.140 & 0.966 $\pm$ 0.005 \\
AV205 & 00:58:23.21223 &  -72:21:35.0756 &  0.110 & 0.959 $\pm$ 0.006 \\
AV211 & 00:58:41.22897 &  -72:26:15.4352 &  0.110 & 0.983 $\pm$ 0.004 \\
AV270 & 01:01:16.99829 &  -72:17:31.1360 &  0.030 & 0.912 $\pm$ 0.004 \\
AV273 & 01:01:27.43692 &  -72:07:06.1109 &  0.060 & 0.884 $\pm$ 0.017 \\
AV297 & 01:02:09.80582 &  -72:00:23.1065 &  0.040 & 0.913 $\pm$ 0.004 \\
AV298 & 01:02:12.31930 &  -72:02:51.5105 &  0.030 & 0.911 $\pm$ 0.015 \\
AV315 & 01:02:49.61001 &  -72:10:14.4805 &  0.060 & 0.922 $\pm$ 0.005 \\
AV338 & 01:03:43.04205 &  -72:15:30.0165 &  0.070 & 0.914 $\pm$ 0.006 \\
AV362 & 01:04:49.34968 &  -72:06:21.7447 &  0.090 & 0.913 $\pm$ 0.005 \\
AV367 & 01:04:52.90948 &  -72:08:36.7848 &  0.070 & 0.916 $\pm$ 0.008 \\
AV382 & 01:05:27.47824 &  -72:48:15.1044 &  0.070 & 0.898 $\pm$ 0.011 \\
AV392 & 01:05:57.93979 &  -71:19:13.6166 &  0.040 & 0.879 $\pm$ 0.016 \\
AV399 & 01:06:08.98986 &  -72:24:08.8972 &  0.040 & 0.906 $\pm$ 0.005 \\
AV443 & 01:09:03.95380 &  -72:32:17.6297 &  0.060 & 0.930 $\pm$ 0.003 \\
AV463 & 01:11:43.16272 &  -72:07:27.5179 &  0.070 & 0.907 $\pm$ 0.012 \\
AV504 & 01:21:48.25650 &  -72:45:59.2387 &  0.030 & 0.850 $\pm$ 0.010 \\ 
\enddata
\end{deluxetable}


\begin{deluxetable}{ccccc}
\tablecaption{Reddening measurements obtained by Larsen et al. (2000) for the LMC and SMC. The measured $(V-I)$ color of the red clump is reported in the fifth column.}
\tablehead{
\colhead{ID} & \colhead{R.A.} & \colhead{Decl.} &    \colhead{E$(B-V)$} & \colhead{$(V-I)_{RC}$}  
}
\startdata
\multicolumn{5}{c}{LMC} \\
\hline
HV982  &  05:29:53  &  -69:09:23 & 0.140 &  1.042 $\pm$ 0.003  \\
\hline
\multicolumn{5}{c}{SMC} \\
\hline
HV1433  & 00:47:11 & -73:41:18 & 0.124 &  0.936 $\pm$ 0.002  \\
HV11284 & 00:49:43 & -72:51:10 & 0.140 & 0.991 $\pm$ 0.008   \\
\enddata
\end{deluxetable}


\begin{deluxetable}{c|cc}
\tablecaption{Unreddened, intrinsic $(V-I)_0$ color of the red clump obtained with different methods of reddening measurements in the LMC. In the last column we report the spread of the $(V-I)_0$ color within each method. }
\tablehead{
\colhead{Method \& reference} & \colhead{$(V-I)_0$} & \colhead{standard deviation} 
}
\startdata
eclipsing binaries Na I (Graczyk et al. 2018) &  0.854 & 0.028 \\
eclipsing binaries Atm. (Graczyk et al. 2018) &  0.818  & 0.050 \\
blue supergiants (Urbaneja et al. 2017) & 0.826 & 0.034 \\
Str\"{o}mgren photometry (Larsen et al. 2000) & 0.857 & -- \\
\enddata
\end{deluxetable}


\begin{deluxetable}{c|cc}
\tablecaption{Unreddened, intrinsic $(V-I)_0$ color of the red clump obtained with different methods of reddening measurements in the SMC. In the last column we report the spread of the $(V-I)_0$ color within each method.}
\tablehead{
\colhead{Method \& reference} & \colhead{$(V-I)_0$} & \colhead{standard deviation} 
}
\startdata
eclipsing binaries Na I (Graczyk et al. 2014) &  0.831 & 0.022 \\
eclipsing binaries Atm. (Graczyk et al. 2014) & 0.818 & 0.043 \\
blue supergiants (Schiller 2010)  & 0.816 & 0.044 \\
Str\"{o}mgren photometry (Larsen et al. 2000) & 0.789 & -- \\
\enddata
\end{deluxetable}


\begin{deluxetable}{cccccc}
\tablecaption{Example of the format of our maps available online for the LMC. Measurements were performed for 3 $\times$ 3 arcmin fields.}
\tabletypesize{\footnotesize}
\tablehead{
\colhead{R.A. [deg]} & \colhead{Decl. [deg]} & \colhead{E$(B-V)$}  &  \colhead{$(B-V)$}  & \colhead{$\sigma$(V-I)$_{RC}$} &  \colhead{N$_{RC}$} \\
\colhead{} & \colhead{} & \colhead{}  &  \colhead{stat. error}  & \colhead{} &  \colhead{} 
}
\startdata
84.7905399298 &  -70.3921738478 &  0.188 & 0.005 & 0.104 & 315 \\
84.9395356186 &  -70.3921738478 &  0.170 & 0.002 & 0.069 & 280 \\
85.0885313073 &  -70.3921738478 &  0.172 & 0.003 & 0.067 & 263 \\
85.2375269961 &  -70.3921738478 &  0.148 & 0.005 & 0.077 & 182 \\
85.3865226849 &  -70.3921738478 &  0.155 & 0.006 & 0.099 & 304 \\
85.5355183736 &  -70.3921738478 &  0.190 & 0.004 & 0.080 & 214 \\
85.6845140624 &  -70.3921738478 &  0.240 & 0.005 & 0.122 & 309 \\
85.8335097511 &  -70.3921738478 &  0.253 & 0.006 & 0.110 & 293 \\
85.9825054399 &  -70.3921738478 &  0.243 & 0.006 & 0.104 & 225 \\
86.1315011287 &  -70.3921738478 &  0.212 & 0.009 & 0.103 & 136 \\
86.2804968174 &  -70.3921738478 &  0.177 & 0.004 & 0.079 & 191 \\
86.4294925062 &  -70.3921738478 &  0.174 & 0.005 & 0.091 & 215 \\
86.578488195  &  -70.3921738478 &  0.150 & 0.004 & 0.087 & 256 \\
86.7274838837 &  -70.3921738478 &  0.191 & 0.006 & 0.079 & 109 \\
86.8764795725 &  -70.3921738478 &  0.191 & 0.005 & 0.071 & 109 \\
87.0254752613 &  -70.3921738478 &  0.170 & 0.006 & 0.088 & 157 \\
87.17447095   &  -70.3921738478 &  0.142 & 0.004 & 0.068 & 177 \\
87.3234666388 &  -70.3921738478 &  0.134 & 0.003 & 0.058 & 177 \\
87.4724623275 &  -70.3921738478 &  0.118 & 0.002 & 0.050 & 234 \\
87.6214580163 &  -70.3921738478 &  0.112 & 0.003 & 0.041 & 57 \\
\enddata
\end{deluxetable}


\begin{deluxetable}{cccccc}
\tablecaption{Example of the format of our maps available online for the SMC. Measurements were performed for 3 $\times$ 3 arcmin fields.}
\tabletypesize{\footnotesize}
\tablehead{
\colhead{R.A. [deg]} & \colhead{Decl. [deg]} & \colhead{E$(B-V)$}  &  \colhead{$(B-V)$}  & \colhead{$\sigma$(V-I)$_{RC}$} &  \colhead{N$_{RC}$} \\
\colhead{} & \colhead{} & \colhead{}  &  \colhead{stat. error}  & \colhead{} &  \colhead{} 
}
\startdata
15.1655554897 &  -72.7497333333 &  0.085 & 0.003 & 0.050 & 152 \\
15.3341633841 &  -72.7497333333 &  0.076 & 0.003 & 0.054 & 160 \\
15.5027712785 &  -72.7497333333 &  0.075 & 0.007 & 0.080 & 364 \\
12.9514471205 &  -72.6997333333 &  0.142 & 0.007 & 0.081 & 144 \\
13.1195825523 &  -72.6997333333 &  0.129 & 0.005 & 0.080 & 236 \\
13.2877179841 &  -72.6997333333 &  0.113 & 0.004 & 0.065 & 188 \\
13.455853416 &  -72.6997333333 &  0.104 & 0.004 & 0.066 & 268 \\
13.6239888478 &  -72.6997333333 &  0.078 & 0.019 & 0.097 & 347 \\
13.7921242796 &  -72.6997333333 &  0.094 & 0.003 & 0.061 & 239 \\
13.9602597114 &  -72.6997333333 &  0.110 & 0.003 & 0.070 & 242 \\
14.1283951432 &  -72.6997333333 &  0.118 & 0.005 & 0.059 & 103 \\
14.296530575 &  -72.6997333333 &  0.102 & 0.006 & 0.077 & 203 \\
14.4646660068 &  -72.6997333333 &  0.113 & 0.004 & 0.068 & 209 \\
14.6328014387 &  -72.6997333333 &  0.094 & 0.003 & 0.065 & 231 \\
14.8009368705 &  -72.6997333333 &  0.095 & 0.005 & 0.045 & 47 \\
14.9690723023 &  -72.6997333333 &  0.115 & 0.005 & 0.069 & 147 \\
15.1372077341 &  -72.6997333333 &  0.100 & 0.007 & 0.077 & 167 \\
15.3053431659 &  -72.6997333333 &  0.089 & 0.004 & 0.068 & 199 \\
15.4734785977 &  -72.6997333333 &  0.088 & 0.004 & 0.066 & 275 \\
15.6416140296 &  -72.6997333333 &  0.078 & 0.013 & 0.082 & 194 \\

\enddata
\end{deluxetable}


\begin{deluxetable}{ccc}
\tablecaption{Differences of the reddening obtained for different types of tracers and reddening obtained in this paper in the LMC. For the RR Lyrae stars, Cepheids and red clump stars the average value of the difference is reported.}
\tablehead{
\colhead{Reddening tracer type} & \colhead{difference} & Comments \\
\colhead{} & \colhead{E$(B-V)$ [mag]} & \colhead{}  
}
\startdata
eclipsing binaries Na I & 0.008 & Graczyk et al. 2018  \\
eclipsing binaries Atm.  &  -0.019 & Graczyk et al. 2018 \\
blue supergiants &  -0.007 & Urbaneja et al. 2017 \\
Str\"{o}mgren photometry  & 0.01  & Larsen et al. 2000 \\
 \hline
BLMC01 &  0.024 & Taormina et al. 2019 \\
BLMC02 &  0.002 & Taormina et al. 2019 \\
HV2274  & 0.005 & Groenewegen \& Salaris 2001 \\
LMC-SC1-105  & 0.015 & Bonanos et al. 2011 \\
NGC 1850  & -0.01 & Walker 1993 \\
NGC 1835  & -0.02 & Lee 1995 \\
RR Lyr  & $\langle$-0.024$\rangle$ & Pejcha \& Stanek 2009 \\
Cepheids  & $\langle$0.004$\rangle$ & Inno et al. 2016  \\
RC  & $\langle$0.061$\rangle$ & Haschke et al. 2011 \\            
RC  & $\langle$0.008$\rangle$ & Choi et al. 2019 \\
\enddata
\end{deluxetable}


\begin{deluxetable}{c|cc|cc}
\tablecaption{Unreddened, intrinsic $(V-I)_0$ color of the red clump obtained with different reddening law R$_V$ values for the LMC with corresponding spread (rms) of the color for each method of reddening measurement.}
\tablehead{
\colhead{} & \multicolumn2c{R$_V$=2.7} &   \multicolumn2c{R$_V$=4.5}  \\
\colhead{Method} & \colhead{$(V-I)_0$} & \colhead{rms} & \colhead{$(V-I)_0$} & \colhead{rms}
}
\startdata
eclipsing binaries Na I &  0.864 & 0.027 &  0.817 & 0.030 \\
eclipsing binaries Atm.  & 0.831 & 0.045 &  0.773 & 0.057 \\
blue supergiants & 0.867 & 0.045 &  0.819 & 0.059 \\
Str\"{o}mgren photometry  & 0.885 & --  &  0.827 & -- \\
\hline
Mean of all methods  & 0.861 & 0.019 & 0.809 & 0.021 \\
\enddata
\end{deluxetable}


\end{document}